\documentclass[lettersize,journal]{IEEEtran}

\usepackage{amsmath,amsfonts,amssymb}
\usepackage{array}
\usepackage{graphicx}
\usepackage{booktabs}
\usepackage{multirow}
\usepackage{threeparttable}
\usepackage{textcomp}
\usepackage{stfloats}
\usepackage{url}
\usepackage{verbatim}
\usepackage{placeins}
\usepackage{cite}
\usepackage{microtype}
\usepackage{soul,xcolor}
\usepackage{adjustbox}
\usepackage{algorithm}
\usepackage{algpseudocode}
\usepackage{makecell}
\usepackage{ragged2e}
\usepackage[hidelinks]{hyperref}

\newif\ifdraftplaceholders
\draftplaceholderstrue

\newcommand{\paperfigure}[6]{%
  \begin{figure}[!t]
    \centering
    \IfFileExists{#1}{\includegraphics[width=#2\linewidth]{#1}}{%
      \fbox{\parbox[c][#3][c]{0.92\linewidth}{\centering
        \textbf{Figure placeholder}\\[2pt]\texttt{\detokenize{#1}}\\[2pt]
        \emph{#4}}}}
    \caption{#5}\label{#6}
  \end{figure}
}

\newcommand{\paperfigurewide}[6]{%
  \begin{figure*}[!t]
    \centering
    \IfFileExists{#1}{\includegraphics[width=#2\textwidth]{#1}}{%
      \fbox{\parbox[c][#3][c]{0.96\textwidth}{\centering
        \textbf{Figure placeholder}\\[2pt]\texttt{\detokenize{#1}}\\[2pt]
        \emph{#4}}}}
    \caption{#5}\label{#6}
  \end{figure*}
}

\newcommand{\NR}{N/R}

\definecolor{darkboundary}{HTML}{50399b}
\definecolor{innerlighter}{HTML}{efeafa}

\begin{document}

\title{VQ-LIC: Shared Vector-Quantized Learned Image Compression on a Resource-Constrained FPGA}

\author{
\IEEEauthorblockN{
Muhammad Fahd Ibrahim Bhatti$^{*}$,
Abdullah Bin Faisal$^{*}$,
Ahsan Usman,\\
Naveed Anwar Bhatti, and
Muhammad Ali Siddiqi
}
\thanks{$^{*}$These authors contributed equally to this work. All authors are with the School of Science and Engineering (SBASSE), Lahore University of Management Sciences (LUMS), Lahore, Pakistan. The machine-learning training and evaluation code, the latency model, and the RTL source code are publicly available at \href{https://github.com/abdullahbinfaisal/VQ-LIC}{\texttt{github.com/abdullahbinfaisal/VQ-LIC}}.}
}

\markboth{}%
{VQ-LIC: Shared Vector-Quantized Learned Image Compression on a Resource-Constrained FPGA}

\maketitle

\begin{abstract}
Learned image compression (LIC) is hard to deploy on severely resource-constrained FPGAs, since how fast it actually runs depends not just on arithmetic count, but also on memory traffic, imbalance between different operations, and how the hardware batches its work. We present VQ-LIC, an asymmetric edge-cloud codec in which a compact INT8 depthwise (DW)-pointwise (PW) analysis transform and multi-codebook vector quantization (VQ) run at the edge on a reusable DW/PW engine pair, while reconstruction is handled by a larger cloud decoder. Since VQ codeword matching is expressible as a dot product, it is mapped directly onto the same PW engine, removing the need for a separate VQ compute array, to our knowledge a first for FPGA LIC. A novel latency model, derived from deterministic RTL cycle counts of an FPGA's read, DW, PW, and write costs, predicts an analysis transform's per-block latency; since VQ shares the same PW datapath, the model applies to VQ as well. Validated directly against silicon, the model predicts deployed analysis and VQ latency within 0.26\% and 0.05\%, and guides the selection of a three-block $16$-$48$-$64$ transform. Post-training codebook reduction then cuts VQ arithmetic and codebook storage by $4\times$ and shrinks the fixed-width latent representation. On a 220-DSP Zynq-7020, VQ-LIC's mid-rate preset reaches 0.1398 bits per pixel at 28.69 dB PSNR and 13.06 dB MS-SSIM on CLIC~2017, outperforming a similarly sized neural encoder and reaching a rate-distortion range comparable to a codec three orders of magnitude larger. The complete 0.1945-kMAC/pixel analysis-VQ pipeline runs at 47.98 frames per second and 42.84 mJ per frame on silicon, using an order of magnitude fewer DSPs than comparable FPGA LIC accelerators while achieving lower bitrate, higher throughput, and lower energy per frame at a modest PSNR tradeoff.
\end{abstract}

\section{Introduction}
\label{sec:intro}

\IEEEPARstart{V}{isual} sensing increasingly occurs on resource-constrained edge platforms, where transmitting raw imagery can dominate bandwidth and energy. Learned image compression (LIC) replaces hand-designed transforms with neural representations optimized for rate--distortion performance~\cite{balle2018,minnen2018}, but recent gains increasingly rely on larger transforms, hyperpriors, autoregressive context models, and attention mechanisms. Their arithmetic, storage, activation-traffic, and entropy-model costs make GPU-oriented codecs difficult to deploy on low-cost embedded devices even after numerical quantization.

FPGA accelerators have addressed this deployment gap through DPU (deep-learning processing unit) mapping~\cite{jia2022}, fine-grained pipelining~\cite{sun2022flic}, parameterized architectures~\cite{chen2025}, algorithm--architecture co-optimization~\cite{sun2024jetcas,sun2025aspdac}, and stream-oriented redesign~\cite{zhang2025streamlic}. These systems establish the value of hardware-aware LIC design but target substantially larger resource envelopes than severely resource-constrained platforms with limited multiplier counts and narrow external-memory interfaces. On such platforms, simply scaling down a larger accelerator is insufficient: arithmetic throughput, operator-specific streaming behavior, external-memory traffic, and finite output batching can each become the bottleneck, so MAC count alone does not reliably predict latency. The challenge is therefore to fit the complete edge encoding pipeline within a tightly limited hardware budget, which raises a central question: \emph{can the latent representation be chosen so that encoding and quantization share the same scarce compute hardware?}

VQ-LIC addresses this question with an asymmetric edge--cloud codec designed for a severely resource-constrained FPGA: the edge device only compresses, while a much larger model in the cloud handles reconstruction. The encoder compresses the image into a compact latent representation and applies vector quantization (VQ), matching pieces of the latent against small, predefined sets of learned patterns, or codebooks. The key insight is that the latent representation is chosen so that this matching step \textit{reuses} the encoder's own multiplier hardware, eliminating the need for a separate VQ compute engine. The resulting discrete indices are compressed with a rANS entropy coder using fixed, pre-shared probability tables rather than a learned hyperprior or autoregressive model, avoiding input-dependent entropy-model inference on the edge.

Each encoder stage uses depthwise filtering (per-channel) followed by pointwise mixing (across channels); the final VQ stage reuses the pointwise mixing hardware. Beyond reuse, latency depends on balancing computation and memory traffic: a stage's latency is set by whichever of reading, filtering, mixing, or writing is slowest, so fewer operations do not guarantee higher speed. We therefore guide the design using a \textit{latency model} derived from the accelerator's own behavior rather than from operation counts and validated against hardware measurements. The model is used first to size the accelerator, then to select the encoder, and finally to guide VQ design, all without implementing each candidate in hardware.

The principal contributions of this work are:

\begin{enumerate}

\item \textbf{A discrete-latent edge--cloud LIC architecture for severely resource-constrained encoding.} A compact INT8 depthwise--pointwise transform, at only 4,491 weights and 194.5 MAC/pixel, and multi-codebook VQ run on the FPGA's programmable logic (PL). The resulting indices are compressed using a table-driven rANS coder on the processing system (PS), removing learned hyperprior and autoregressive entropy-model inference from the edge path while relocating reconstruction capacity to the cloud. The complete edge pipeline sustains 47.98~frames/s at 720p on the 220-DSP Zynq-7020.

\item \textbf{A novel compute--communication latency model for resource-aware accelerator provisioning.} Competing read, depthwise, pointwise, and write costs are modeled from tensor geometry and hardware parallelism, provisioning resources by bottleneck rather than MAC share. Verified over more than 100 RTL configurations and against silicon, the model's per-block latency predictions agree within 0.3\% with no fitted correction term, correctly identifying the bottleneck in every deployed block. Applied before training to candidate tensor geometries without implementing any of them in hardware, it guides selection of the $16$--$48$--$64$ encoder, cutting predicted latency by 24.0\% relative to the six-block reference while improving proxy reconstruction quality.

\item \textbf{Reuse of the validated pointwise engine for codeword scoring.} Because VQ codeword scoring is expressible as pointwise MACs, it is mapped directly onto the analysis transform's existing pointwise MAC array rather than a dedicated VQ compute engine, letting VQ run on already-instantiated hardware, to the best of our knowledge a first for FPGA learned image compression.

\end{enumerate}

Section~\ref{sec:related} reviews LIC and FPGA-oriented co-design and positions VQ-LIC against the closest prior systems. Sections~\ref{sec:methodology} and~\ref{sec:arch} present the codec and the compute--communication model and accelerator, Section~\ref{sec:method} defines the encoder-exploration and measurement methodology, and Section~\ref{sec:results} reports results and compares with prior FPGA LIC implementations. Section~\ref{sec:conclusion} concludes the discussion.

\section{Related Work}
\label{sec:related}

\subsection{Discrete-Latent and Efficient Learned Image Compression}
\label{sec:related-lic}

Learned image compression replaces hand-designed transforms and probability models with neural components optimized for rate--distortion performance~\cite{balle2017}. Transformer-based extensions can further improve compression efficiency, but at increased computational and memory cost~\cite{liu2023tcm,duan2023attention}. The dominant approach rounds a continuous latent representation to integers (scalar quantization) and entropy-codes it using a learned probability model~\cite{balle2018,minnen2018}.

VQ-VAE~\cite{oord2017} established a learned codebook of representations, allowing codebook indices to be transmitted instead of latent values. Later compression work built on this with entropy-aware soft-to-hard VQ~\cite{agustsson2017softvq}, probabilistic multi-codebook quantization~\cite{zhu2022mcquic}, product-quantized representations with learned entropy modeling~\cite{elnouby2023pqmim}, nonlinear and multi-stage vector transform coding~\cite{feng2023nvtc}, and quantized hierarchical latent models~\cite{duan2023qres}.

Separately, efficiency work has focused on reduced numerical precision through channel splitting~\cite{new:qlic} and rate--distortion-aware post-training quantization~\cite{shi2023ptqlic}, and on reduced model capacity through slimmable configurations~\cite{yang2021} or distillation~\cite{mazouz2025}. Asymmetric designs such as MCUCoder~\cite{2025mcucoder} shrink the encoder for edge deployment and keep a larger decoder. To our knowledge, however, such compact encoders have not been mapped onto multiplier-constrained FPGAs.

\subsection{FPGA Acceleration of LIC}
\label{sec:related-hw}

FPGA LIC accelerators span vendor overlays, dedicated pipelines, and explicit model--hardware co-design. FPX-NIC~\cite{jia2022} maps neural coding through the Xilinx DPU, while F-LIC~\cite{sun2022flic} uses a fine-grained interlayer pipeline with flexible channel parallelism and zero skipping. Sun et al.~\cite{sun2024jetcas,sun2025aspdac} further co-optimize network channel dimensions with realizable FPGA parallelism, accounting for routing and activation-buffer constraints that limit usable DSP concurrency across ZC706-, KU115-, and VCU118-class devices.

StreamLIC~\cite{zhang2025streamlic} instead restructures transform and autoregressive processing into a row-streaming architecture, retaining hyperprior and context-model inference while reducing serialization cost. X-LIC~\cite{chen2025} uses a parameterized accelerator to improve utilization across heterogeneous LIC operators, whereas Mazouz et al.~\cite{mazouz2025,mazouz2026tmm} combine model dimensioning, distillation, quantization, and hardware-friendly nonlinear processing for FPGA deployment. Collectively, these works adapt conventional learned-codec pipelines to the target hardware through dataflow, accelerator, or network-structure optimization, while predominantly retaining scalar-quantized transforms with learned entropy processing.

Dedicated vector-quantizer hardware predates its use in learned compression, including full-search VLSI architectures~\cite{park1993} and FPGA designs based on partial-distance search~\cite{pds2006}. These implement nearest-neighbor search as a dedicated compute array separate from any surrounding convolution or transform hardware, an additional resource cost that a severely multiplier-constrained platform can ill afford.

Beyond LIC-specific designs, depthwise-separable convolution~\cite{howard2017,chollet2017} has also been accelerated on FPGAs~\cite{bai2018}, but reducing arithmetic does not determine how a small fixed multiplier budget should be divided between depthwise streaming and pointwise computation. Under limited external bandwidth, depthwise and pointwise operations place different demands on the hardware, so a good split between them cannot be inferred from MAC share alone.

\subsection{Positioning of VQ-LIC}
\label{sec:related:positioning}

Hardware-aware neural architecture search already establishes that arithmetic complexity is an imperfect proxy for realized cost: MnasNet~\cite{tan2019mnasnet}, ProxylessNAS~\cite{cai2019proxylessnas}, and FBNet~\cite{wu2019fbnet} incorporate target-device performance into architecture selection, while FNAS~\cite{jiang2019fnas} uses an FPGA performance model to avoid implementing every candidate. VQ-LIC therefore does not claim the general observation that MAC count can differ from hardware latency; the distinction is where that observation is derived from and how far it is carried.

The closest FPGA LIC co-designs, Sun et al.~\cite{sun2024jetcas,sun2025aspdac} and StreamLIC~\cite{zhang2025streamlic}, improve hardware efficiency by adapting network structure and dataflow to the target FPGA. VQ-LIC instead derives its hardware model by deterministic cycle counting of the realized accelerator's behavior, rather than fitting it to observed latency. This same model is then carried further: it guides the choice of encoder architecture, resolving depthwise--pointwise resource allocation. Because VQ codeword scoring reuses the same pointwise engine rather than a dedicated compute array, the latency model also applies directly to VQ, allowing the hardware cost of post-training codebook-complexity reduction to be evaluated without a separate quantizer-specific model.

\section{VQ-LIC Codec and Training Methodology}
\label{sec:methodology}

\begin{figure}[t]
\centering
\includegraphics[width=1\linewidth]{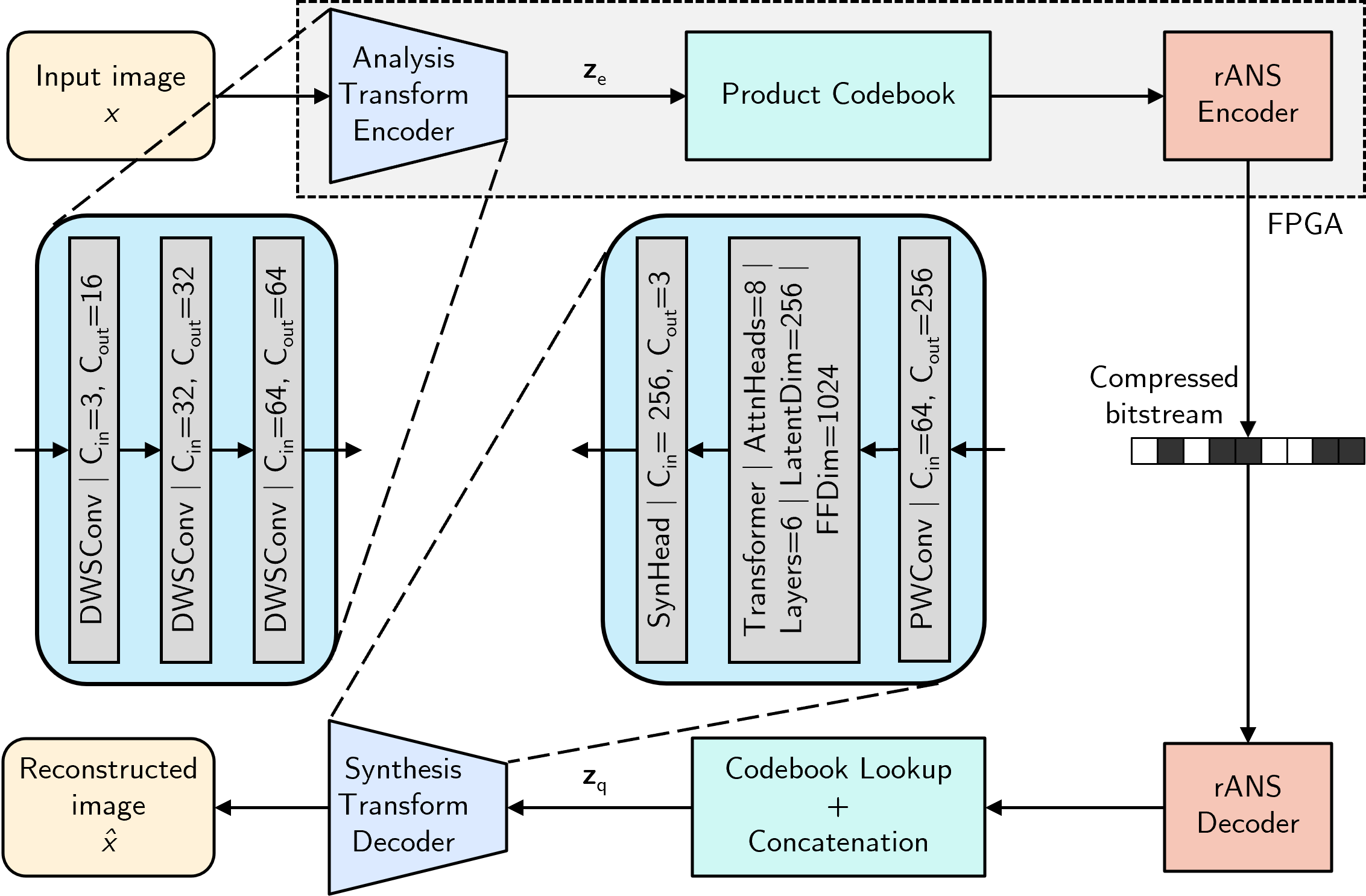}
\caption{End-to-end VQ-LIC codec. DWSConv: depthwise-separable convolution (depthwise followed by pointwise).}
\label{fig:codec}
\end{figure}

Fig.~\ref{fig:codec} shows the VQ-LIC codec: a compact INT8 depthwise--pointwise analysis transform and multi-codebook VQ on the edge, table-driven rANS coding, and a larger cloud-side decoder.

\subsection{Analysis Transform}
\label{sec:methodology:encoder}

For a $1280\times720$ (720p) input image $\mathbf{x}$, the selected analysis transform $f_\theta$, parameterized by encoder weights $\theta$, produces the continuous latent $\mathbf{z}_e\in\mathbb{R}^{90\times160\times64}$ using three stride-2 depthwise--pointwise blocks with channel schedule $C^\star=(16,48,64)$. The transform contains 4,491 convolutional weights; each block applies a $3\times3$ depthwise convolution followed by a $1\times1$ pointwise convolution, with batch normalization folded into the preceding convolution for deployment. Intermediate activations are requantized to INT8, and the final 64-channel latent is stored in unsigned INT8, avoiding an additional conversion before quantization. The analysis transform executes in the programmable logic (PL).

\subsection{Hardware-Aware Product-Codebook Vector Quantization}
\label{sec:methodology:vq}

\begin{figure}[t]
\centering
\includegraphics[width=1\linewidth]{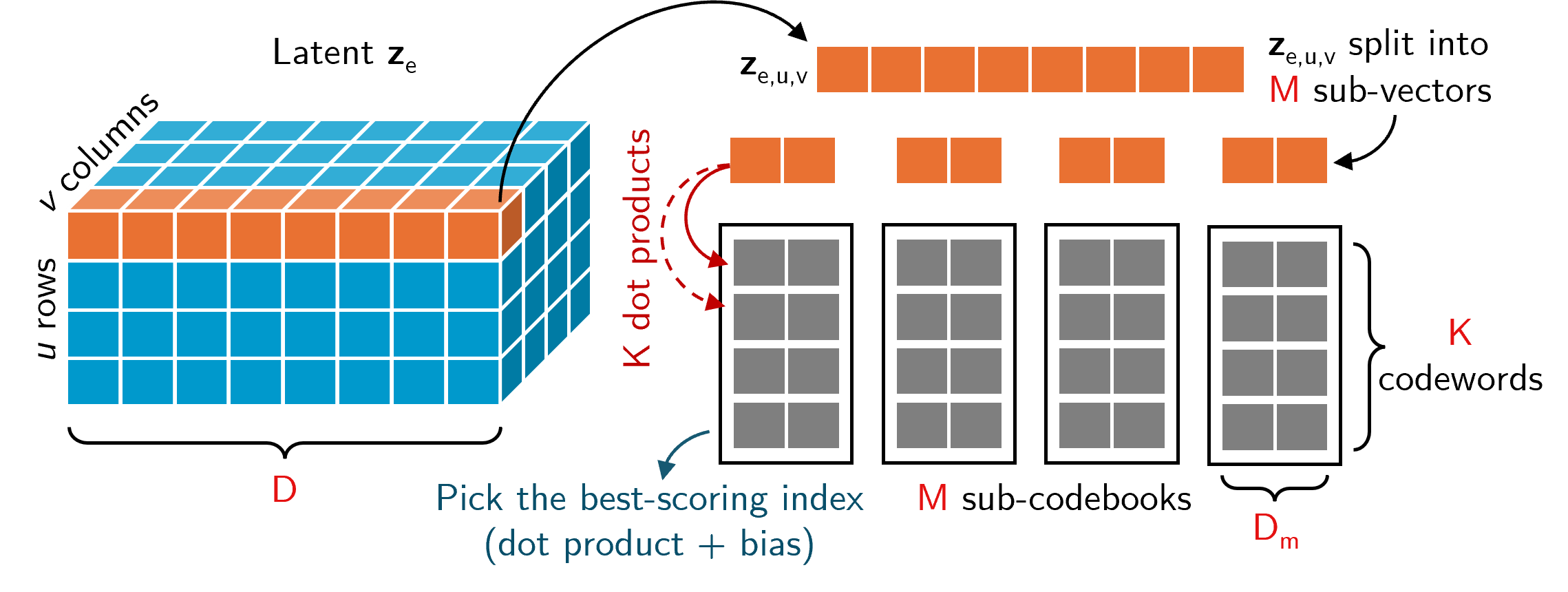}
\caption{Multi-codebook VQ on a $D$-dimensional latent vector $\mathbf{z}_{e,u,v}$ at spatial position $(u,v)$. The vector is split into $M$ disjoint sub-vectors of dimension $D_m=D/M$, and each sub-vector is scored against its own $K$-entry sub-codebook by computing a dot product with every candidate codeword; the lowest-scoring candidate's index is retained.}
\label{fig:vq}
\end{figure}

For a latent of dimension $D=64$, multi-codebook VQ, executed in the programmable logic (PL), partitions each spatial vector $\mathbf{z}_{e,u,v}$ into $M$ disjoint sub-vectors $\mathbf{z}^{(m)}_{e,u,v}$, each of dimension $D_m=D/M$, and matches each sub-vector to the closest of $K$ candidate codewords in its own codebook $\mathcal{E}^{(m)}=\{\mathbf{e}^{(m)}_j\}_{j=0}^{K-1}$ (see Fig.~\ref{fig:vq}). Rather than always choosing the closest codeword, we additionally bias this assignment toward codewords that are cheaper to entropy-code, using entropy-constrained vector quantization (ECVQ)~\cite{1989ecvq} applied independently within each sub-codebook. The rate-aware codeword index assigned to latent position $(u,v)$ and sub-codebook $m$ is
\begin{equation} k_{u,v,m}=\arg\min_j\left[\left\|\mathbf{z}_{e,u,v}^{(m)}-\mathbf{e}_{j}^{(m)}\right\|_2^2-\beta_{\mathrm{rate}}\log_2p_{j}^{(m)}\right], \label{eq:ecvq_assignment} \end{equation}
where $p_j^{(m)}$ is an exponential-moving-average (EMA) estimate of the usage probability of codeword $j$ in sub-codebook $m$, and $\beta_{\mathrm{rate}}$ controls the rate pressure.

The quantizer dimensions are fixed before co-design, targeting a $0.1\!-\!0.5$~bpp regime that covers the FPGA LIC operating points in Table~\ref{tab:priorwork}. The three stride-2 analysis blocks downsample by $\prod_j s_j=8$, giving a $D=64$ latent; with $M=4$ sub-vectors of dimension $D_m=16$ and a reference codebook of $K_0=256$ entries, the uncoded rate is $M\lceil\log_2K_0\rceil/\left(\prod_j s_j\right)^2=32/64=0.5$~bpp, the upper end of the target range. A downsampling of 16 would instead give 0.125~bpp, and 4 would give 2~bpp with four times as many latent positions. This 32-bit index budget also sets a practical product-quantization balance for $M$~\cite{jegou2011}: fewer sub-codebooks sharply increase codebook size and search cost ($M=2$ needs $K=65{,}536$, or $256\times$ more multiplies per latent position), while more sub-codebooks shrink $D_m$ toward scalar quantization.

Encoder exploration uses the capacity-rich reference quantizer with initial configuration $(M_0,K_0,D_{m,0})=(4,256,16)$. After topology selection, only the codebook cardinality $K$ is refined post hoc, with $M$ and $D_m$ held fixed. For each candidate $K\in\{256,128,96,64,32\}$, entries within each sub-codebook's original $K_0=256$-entry pool are ranked by their EMA training-usage prior, and only the $K$ highest-prior entries are retained; evaluation-set assignments are not used to choose the retained codes. The surviving prior is renormalized, and the resulting codec is evaluated jointly on reconstruction quality, entropy-coded rate, codebook storage, and modeled shared-engine VQ latency. This procedure selects the final configuration $(M^\star,K^\star,D_m^\star)=(4,64,16)$. Because refinement only ever reduces $K$ from $K_0$, every deployed configuration remains within the $0.5$~bpp ceiling set by $K_0$; $\beta_{\mathrm{rate}}$ then controls the entropy-coded operating points within it (Section~\ref{sec:methodology:decoder}).

The selected quantizer therefore represents each latent position using four 6-bit indices, or 24 uncoded bits per position. The resulting discrete indices are entropy-coded on the Cortex-A9 processing system (PS) using rANS with pre-shared conditional probability tables. Because the transform downsamples spatially by $8\times8$, the uncoded representation corresponds to a fixed-width rate of $24/64=0.375$ bits per pixel (bpp) before entropy coding; the transmitted rate is instead measured from the final rANS payload. For integer deployment, let $\tilde{\mathbf{z}}_{e,u,v}^{(m)}$ denote the unsigned-INT8 representation of encoder sub-vector $\mathbf{z}_{e,u,v}^{(m)}$, and let $\tilde{\mathbf{e}}_{j}^{(m)}$ denote the unsigned-INT8 representation of codeword $\mathbf{e}_{j}^{(m)}$ in the same domain with zero point $z_0$. Expanding the rate-aware squared-distance objective in Eq.~\eqref{eq:ecvq_assignment} and dropping the query-norm contribution common to all candidate codewords gives the deployed assignment index
\begin{equation}
\begin{aligned}
k^\star_{u,v,m}
&=
\arg\min_j
\Bigg[
\left\|
\tilde{\mathbf{e}}_{j}^{(m)}-z_0
\right\|_2^2
-
\tilde{\beta}_{\mathrm{rate}}\log_2 p_j^{(m)}
\\
&\qquad
-
2
\left(
\tilde{\mathbf{z}}_{e,u,v}^{(m)}-z_0
\right)^T
\left(
\tilde{\mathbf{e}}_{j}^{(m)}-z_0
\right)
\Bigg].
\end{aligned}
\label{eq:vq_pw_score}
\end{equation}
Here, $\tilde{\beta}_{\mathrm{rate}}$ denotes the rate weight after deployment scaling. The codeword-norm and prior-dependent terms are precomputed for each codeword, while the dot-product term is evaluated by the existing pointwise MAC array; bias addition and the comparison of resulting scores to find the lowest-scoring codeword use auxiliary logic detailed in Section~\ref{sec:arch}. During floating-point training, codebooks are updated by exponential-moving-average statistics, and gradients are passed to the encoder through a straight-through estimator.

\subsection{Cloud Decoder and Training Objective}
\label{sec:methodology:decoder}

At the reconstruction endpoint, decoded indices recover the quantized latent $\mathbf{z}_q$, and the cloud decoder $g_\phi$, parameterized by weights $\phi$, produces the reconstructed image $\hat{\mathbf{x}}=g_\phi(\mathbf{z}_q)$. It contains approximately 6.51 million parameters and lies outside the Zynq-7020 hardware boundary. It first projects the quantized latent to a 256-dimensional token representation and processes it using six Transformer layers with eight attention heads and a feed-forward dimension of 1024. Self-attention is restricted to fixed $14\times14$-token windows, with the window partition shifted by seven tokens on alternating layers to exchange information across adjacent windows. A periodic two-dimensional sinusoidal positional encoding with the same $14\times14$ period ensures that the positional vocabulary is unchanged as image resolution increases. The complete latent grid is decoded in a single forward pass; windowing applies only within the self-attention operator, and the image itself is not spatially tiled or overlap-blended. After the Transformer, a convolutional synthesis head with three sub-pixel upsampling stages reconstructs the RGB image. The final Tanh output corresponds to the $[-1,1]$ image representation used during training.

The encoder, VQ codebooks, and decoder are optimized jointly during floating-point training using an objective that combines the structural similarity index measure (SSIM), $\ell_1$ reconstruction error, and latent commitment:
\begin{equation}
\begin{aligned}
\mathcal{L}_{\mathrm{train}}
={}&
\alpha
\left(
1-\operatorname{SSIM}(\mathbf{x},\hat{\mathbf{x}})
\right)
+
(1-\alpha)
\left\|
\mathbf{x}-\hat{\mathbf{x}}
\right\|_1\\
&+
\beta
\left\|
\mathbf{z}_{e}
-
\operatorname{sg}[\mathbf{z}_{q}]
\right\|_2^2,
\end{aligned}
\label{eq:total_loss}
\end{equation}
where $\alpha=0.84$ is the reconstruction weight balancing the SSIM and $\ell_1$ terms, following the setting reported to work best for this loss combination in~\cite{zhaoalpha}, $\beta=0.25$ is the commitment weight, and $\operatorname{sg}[\cdot]$ denotes stop gradient. Codebook vectors are updated through EMA rather than a separate gradient-based embedding loss.

During training, the rate-aware assignment of Eq.~\eqref{eq:ecvq_assignment} is used to obtain the codec's operating points. The final low-rate preset, mid-rate preset, and high-rate preset are three independently trained models that use $\beta_{\mathrm{rate}}=1.0$, $0.4$, and $0.3$, respectively. During encoder exploration all candidates instead use the common $(4,256,16)$ reference quantizer with $\beta_{\mathrm{rate}}=0$, so the VQ rate term does not affect their hardware-latency ordering; post-training codebook refinement is performed only after selecting $C^\star$, as described in Section~\ref{sec:method:vq_sweep}.

\subsection{Floating-Point and Quantization-Aware Training}
\label{sec:methodology:training}
Models are trained on Open Images V7~\cite{kuznetsova2020openimages}, a dataset of 1.7 million images, using $224\times224$ randomly augmented crops normalized to $[-1,1]$. After architecture selection, the encoder, VQ codebooks, and cloud decoder are trained jointly in FP32 for 2 epochs using AdamW with an initial learning rate of $2.8\times10^{-4}$ and a decaying schedule, with batch size 32. The model is subsequently fine-tuned for 2 further epochs using quantization-aware training (QAT) for the encoder, with the codebook frozen during this stage to prevent quantization noise from corrupting the learned codes, while the decoder remains in FP32. Detailed optimizer schedules and framework-specific training settings are provided with the released implementation.

The deployed encoder uses per-channel symmetric signed INT8 weights and per-tensor asymmetric unsigned INT8 activations, with convolution products accumulated at wider integer precision before requantization. Batch normalization is folded into the preceding convolution before deployment. The final stored latent and VQ codebooks use a common quantization domain with zero point $z_0=128$: latent values enter VQ as unsigned INT8, while codewords are stored in the pointwise weight memory as signed zero-point-centered INT8 values. In VQ mode, the existing pointwise MAC datapath evaluates the dot-product term of~\eqref{eq:vq_pw_score}; score formation applies the precomputed codeword-dependent bias to the wide accumulator output before the lowest-scoring codeword is selected, while pointwise requantization, ReLU, and output clamping are bypassed for VQ results. The selected $(M^\star,K^\star)=(4,64)$ quantizer therefore produces four 6-bit indices per latent position.

\section{VQ-LIC Compute--Communication Co-Design and FPGA Architecture}
\label{sec:arch}

\begin{figure}[t]
\centering
\includegraphics[width=1\linewidth]{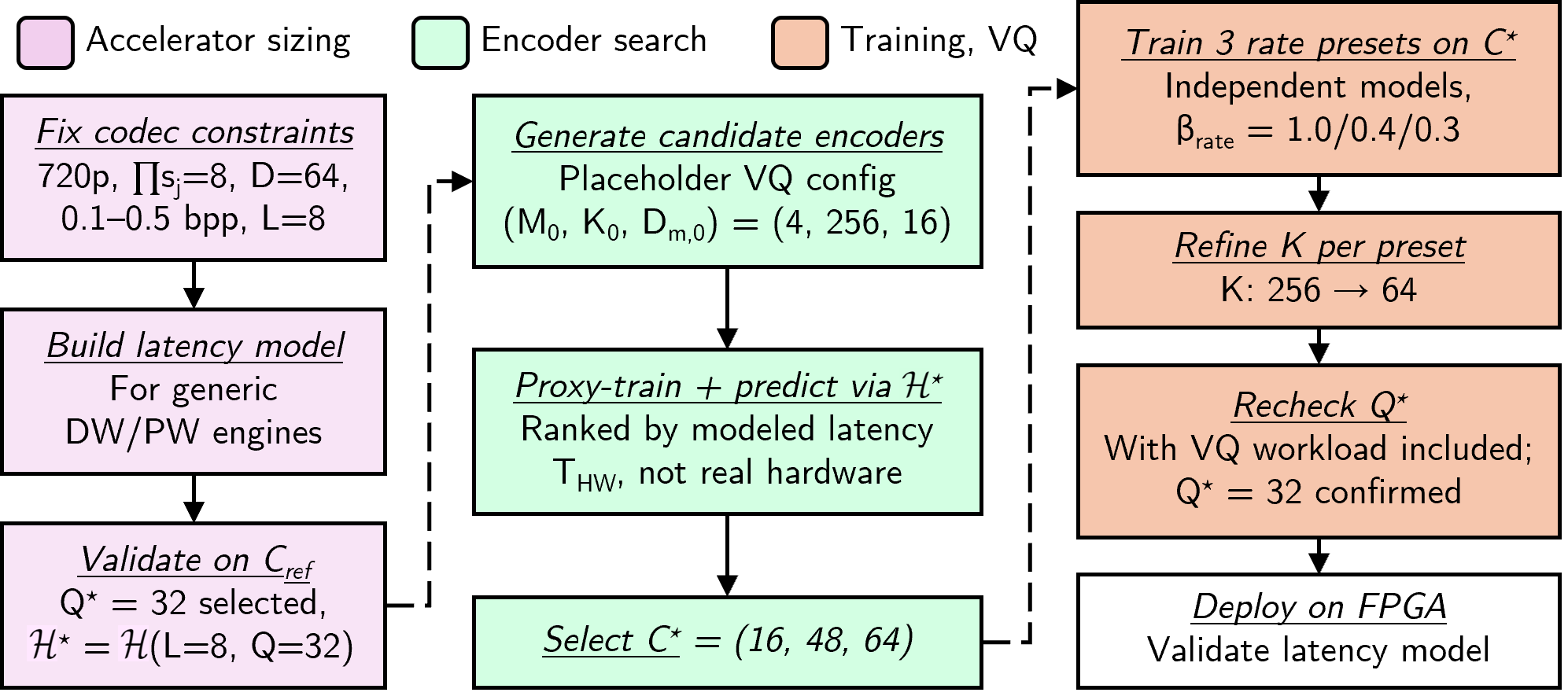}
\caption{Overview of the staged co-design pipeline. A latency model, built from RTL cycle counts and validated on a reference encoder, is used to select the accelerator configuration, then to rank candidate encoders and select the deployed topology, and to guide post-training codebook refinement across the three rate presets, before being validated again on silicon.}
\label{fig:flow}
\end{figure}

VQ-LIC treats accelerator provisioning, encoder topology, and latent quantization as a staged algorithm--architecture co-design problem: on a severely resource-constrained FPGA, minimizing MAC count alone does not necessarily minimize latency, since latency also depends on tensor geometry, operator type, memory traffic, hardware parallelism, and finite batching. We distinguish the encoder topology $C$, the VQ configuration $(M,K,D_m)$, and the reusable accelerator $\mathcal{H}$, and evaluate their effects through reconstruction quality, coded rate, storage, and hardware latency. Rather than a monolithic joint search, we first model the latencies induced by candidate tensor geometries to establish and validate the accelerator operating point, then reuse the same model as a latency oracle for encoder exploration, and finally reuse it again, since codeword scoring shares the same pointwise datapath, to guide post-training codebook-cardinality refinement. Fig.~\ref{fig:flow} summarizes this staged co-design process.

\subsection{Design Constraints and Workload}
\label{sec:arch:problem}

The target platform is the Zynq-7020, which provides 220 DSP48E1 slices and a 64-bit communication path between external DDR memory and the programmable logic (PL). Convolutional weights are represented as signed INT8 values and intermediate activations as unsigned INT8 values, while accumulation is performed at wider integer precision before requantization. The deployed VQ interface uses the same unsigned INT8 latent domain with zero point $z_0=128$, allowing the pointwise MAC array to be reused for centered codeword dot products as described in Section~\ref{sec:arch:organization}.

Within the specification of Section~\ref{sec:methodology:vq}, a candidate encoder is represented by its sequence of output-channel widths, $C=(c_1,c_2,\ldots,c_N)$, where $N=|C|$ denotes the number of depthwise--pointwise blocks and $c_{\mathrm{out},j}$ is the output-channel width of block $j$. The corresponding input-channel count is $c_{\mathrm{in},1}=3$ for the first block, and $c_{\mathrm{in},j}=c_{\mathrm{out},j-1}$ for $j>1$. Since the final block's output feeds the latent, $c_{\mathrm{out},N}=D=64$, and reaching the required $\prod_j s_j=8$ downsampling needs three stride-2 blocks, so $N\ge3$. Our starting point is a six-block encoder, $C_{\mathrm{ref}}=(16,32,32,32,64,64)$, which bounds the explored depth at $N\le6$ and provides the workload on which the accelerator is provisioned (Section~\ref{sec:arch:selection}). Intermediate widths are drawn from $\{16,32,48,64\}$, bounded by $D$.

\begin{figure}[t]
\centering
\includegraphics[width=1\linewidth]{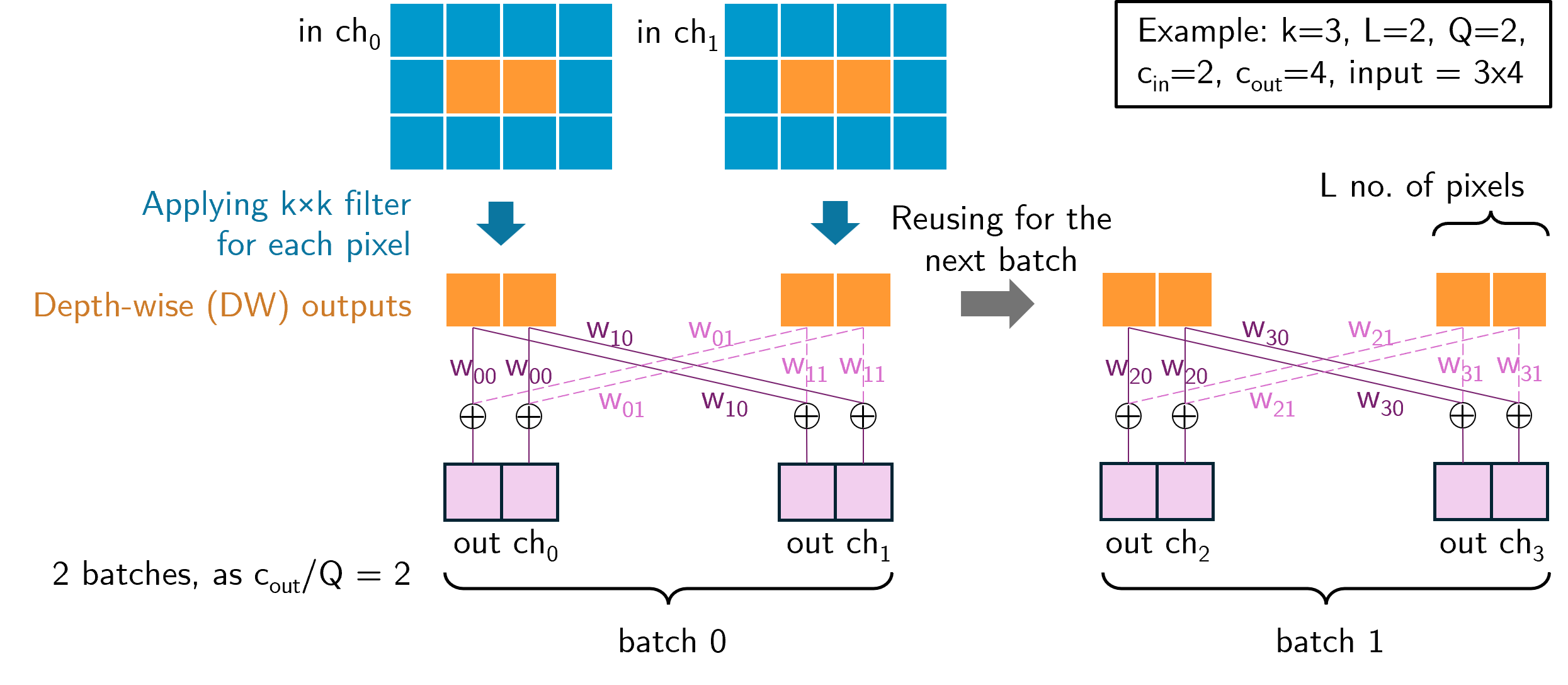}
\caption{Depthwise--pointwise datapath on a small example ($L=2$, $Q=2$, $c_{\mathrm{in}}=2$, $c_{\mathrm{out}}=4$, $3\times3$ filter, $3\times4$ input). Each of the $L=2$ streamed pixels is filtered per input channel to form the depthwise outputs, which are then combined across input channels using pointwise weights $w_{c_{\mathrm{out}},c_{\mathrm{in}}}$ to form each output channel. Because only $Q=2$ output channels can be evaluated in parallel, the four output channels are produced across $c_{\mathrm{out}}/Q=2$ batches, with the same pointwise datapath reused for the second batch.}
\label{fig:depthwise-pointwise}
\end{figure}

Across the evaluated search space, the leading three stride-2 blocks reduce a 720p input to the common $160\times90$ latent grid; when $N>3$, all subsequent blocks use unit stride and therefore operate at the latent spatial resolution. Consequently, different neural architectures present different tensor geometries to the same reusable accelerator.
For candidate $C$, block $j$ presents the workload descriptor
\begin{equation}
\mathcal{W}_{j}(C)=\left(H_j,W_j,c_{\mathrm{in},j},c_{\mathrm{out},j},s_j\right),
\label{eq:block_workload_descriptor}
\end{equation}
where $H_j$ and $W_j$ denote the input spatial dimensions of the block. We denote an accelerator configuration by $\mathcal{H}(L,Q)$, where $L$ is the spatial stream width and $Q$ is the number of pointwise output channels evaluated in parallel. The quantities in $\mathcal{W}_{j}(C)$ directly parameterize the external-memory, depthwise, pointwise, and output latencies developed in Section~\ref{sec:arch:model}. Hence, for a fixed accelerator configuration, the neural architecture induces the per-block read, depthwise, pointwise, and write latencies
\begin{equation}
C\longrightarrow\left\{T_{\mathrm{read},j},T_{\mathrm{DW},j},T_{\mathrm{PW},j},T_{\mathrm{write},j}\right\}_{j=1}^{N},
\label{eq:model_to_hardware_mapping}
\end{equation}
defined formally in Section~\ref{sec:arch:model}.
This mapping is the mechanism by which the FPGA model later provides hardware feedback during structured encoder exploration. The implemented 64-bit interface fixes $L=8$ spatial lanes. The pointwise datapath packs two spatial products into each DSP48E1 slice, so a pointwise multiplier array with spatial parallelism $L$ and output-channel parallelism $Q$ requires
\begin{equation}
D_{\mathrm{PW}}(L,Q)=\frac{LQ}{2}
\label{eq:pw_dsp_cost}
\end{equation}
DSP48E1 slices.
Equation~\eqref{eq:pw_dsp_cost} counts only the structural packed pointwise multiplier grid and is not used as a predictor of complete post-synthesis DSP occupancy. Accumulation, requantization, and related arithmetic may be mapped between DSP48E1 and LUT resources by synthesis depending on the surrounding design and available device resources; complete resource feasibility is therefore established from synthesis in addition to the analytical latency model. Depthwise arithmetic width is not independently tunable: once $L$ fixes the spatial lanes, the depthwise datapath is fixed accordingly, while its realized latency additionally depends on the input-stream and sliding-window-generator schedule developed in Section~\ref{sec:arch:model}. Thus, $Q$ is the independently tunable pointwise-parallelism parameter for the implemented datapath, while the learned channel dimensions determine the workload presented to it. Fig.~\ref{fig:depthwise-pointwise} illustrates this depthwise--pointwise datapath and output-channel batching on a small worked example.

\begin{figure}[t]
\centering
\includegraphics[width=0.8\linewidth]{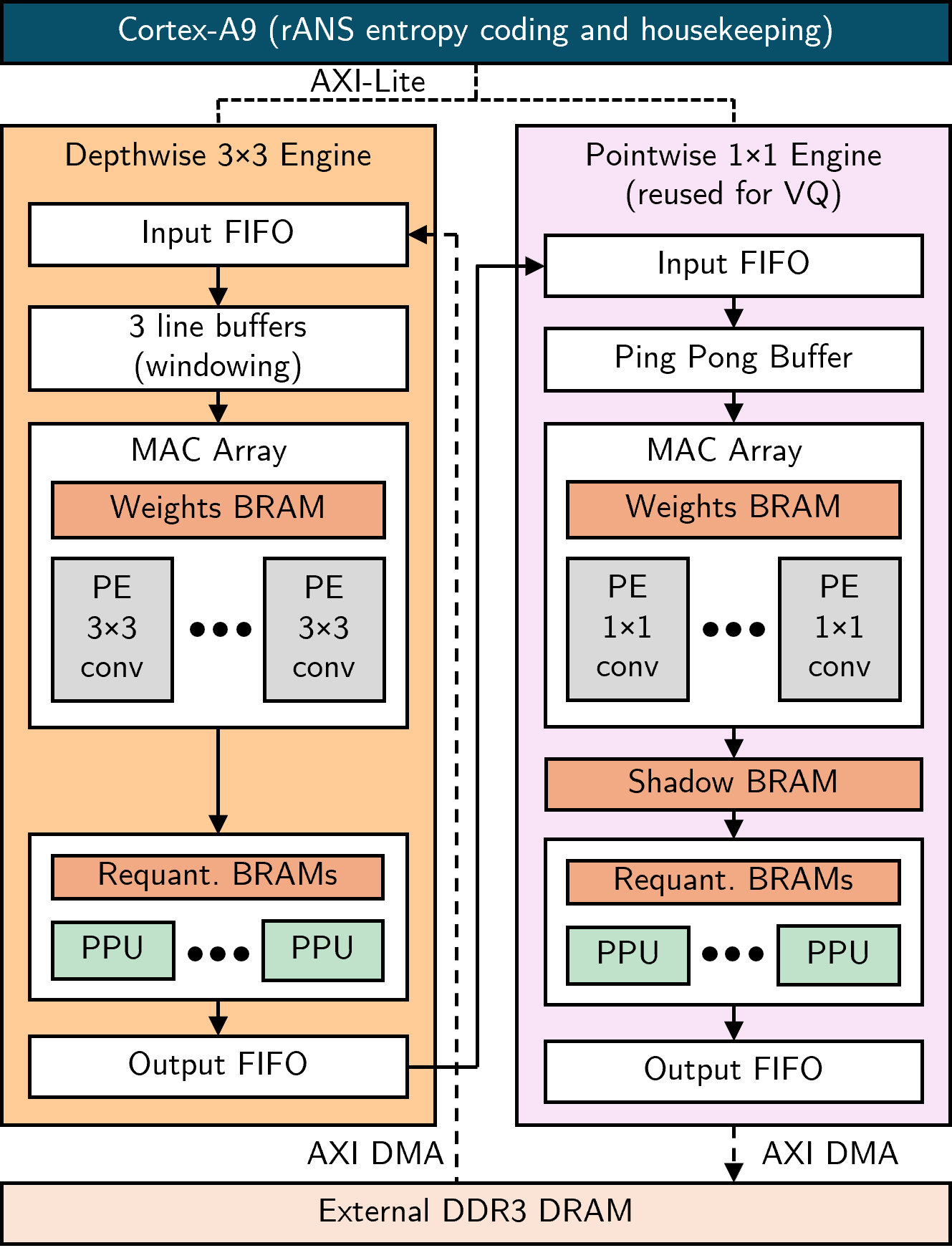}
\caption{System-level organization. The Cortex-A9 configures the reusable accelerator and controls the 64-bit DMA paths between external DDR3 DRAM and the PL. Within each encoder block, the depthwise output streams directly into the pointwise engine; only the pointwise output is materialized in DDR between blocks. A single depthwise--pointwise engine pair is reused sequentially across the encoder blocks. After the final analysis-transform block, VQ is performed in the PL and the resulting discrete indices are transferred to the PS for rANS entropy coding.}
\label{fig:system}
\end{figure}

\subsection{Accelerator Organization}
\label{sec:arch:organization}

The encoder blocks reuse a common depthwise--pointwise engine pair, allowing the two convolution operators to be provisioned independently (see Fig.~\ref{fig:system}). The pointwise engine is additionally reused after analysis for vector quantization, so its output-channel parallelism $Q$ governs the latency of both $1\times1$ convolution and VQ codeword scoring.

\subsubsection{Depthwise Engine}
The depthwise engine parallelizes across spatial positions rather than channels: each cycle, it computes the full $3\times3$ filter for $L$ adjacent pixels within one channel. This avoids underutilizing the engine in the three-channel first block, where channel-wise parallelism would leave most lanes idle, and it naturally matches the $L$-wide pixel stream. Line buffers and a sliding-window generator assemble each $3\times3$ neighborhood from the streamed rows, generating zero-point padding on chip at image edges rather than storing padded data in memory. Wide-precision accumulation results are then requantized to INT8 by an $L$-lane post-processing unit (PPU), which keeps pace with the depthwise engine exactly and packs its output into the same 64-bit activation format the pointwise engine expects, so the two engines connect directly with no reformatting.

\subsubsection{Pointwise Engine}

The pointwise engine is output-stationary: each output accumulator stays fixed while inputs stream through it, so each cycle, the $L$ activations of one input channel are broadcast to $Q$ output-channel banks, which accumulate across all input channels. Input activations are staged through a ping-pong buffer so that one bank can feed the pointwise datapath while the other is refilled for the next activation group. Alternating the two banks overlaps input staging with computation and prevents buffer refill from unnecessarily interrupting the pointwise accumulation schedule. 

Two adjacent, zero-point-corrected activations sharing a common weight are combined into one wide operand and multiplied in a single DSP48E1 operation using standard INT8 DSP packing~\cite{xilinx2017int8}, recovering both exact signed products from the packed result with a small correction for negative operands. Thus, one DSP48E1 recovers two exact signed products, so the selected $L=8,Q=32$ grid evaluates 256 products/cycle using 128 DSP48E1 slices. Because a batch's outputs are not immediately consumed once finalized, they are copied to a second, shadow storage bank at two channels per cycle so that the next batch's accumulation can begin immediately, while the $L$-lane PPU retires (reads out and requantizes to INT8) the preceding group from shadow storage in parallel. When the final batch has fewer than $Q$ real output channels, this \emph{split-shadow} schedule retires only those valid channels rather than the full batch width.

\subsubsection{Shared Pointwise VQ Mode}

VQ reuses the pointwise engine by representing candidate codewords as output channels and storing centered codebook coefficients in the existing INT8 weight memories. For the selected $(M,K,D_m)=(4,64,16)$ quantizer and $Q=32$, each 64-entry sub-codebook needs two passes to evaluate all its codewords, giving eight $Q$-wide batches across the four sub-codebooks, none shared with another sub-codebook. Because each batch belongs to a single sub-codebook, every codeword's $D_m=16$ values fill it completely, so no slots are wasted on zero-padding.

The existing MAC array computes the dot-product term of Eq.~\eqref{eq:vq_pw_score} for each candidate codeword $j$; the remaining, codeword-dependent terms of that same equation are precomputed and simply added to form the assignment score, with only bias storage, score/argmin logic, and index-packing control added around the reused datapath. The running minimum is updated as candidate scores retire.

\subsubsection{Integer Post-Processing}

Both convolution datapaths include an $L$-lane integer PPU that requantizes wide accumulator outputs to the INT8 stream representation, but the two PPUs keep pace differently. The depthwise PPU can accept a new $L$-pixel group every cycle, matching the depthwise engine's own rate, so it never falls behind and adds no steady-state delay of its own. The pointwise PPU, however, finishes only one output channel per cycle, so with $Q$ channels per batch, it takes $Q$ cycles to clear a batch even if the engine could move faster; this delay is why it appears explicitly in the pointwise scheduling model of Section~\ref{sec:arch:model}.

\subsubsection{Depthwise--Pointwise Fusion and External Memory}

Within each encoder block, depthwise outputs feed directly into the pointwise engine on chip rather than being written to and read back from external memory, eliminating that write/read round trip. Outputs between blocks are still written to memory as before, so this saving applies only within a block, not between blocks.

The depthwise engine computes one channel fully before moving to the next, but packages its output spatial position by spatial position across channels, matching what the pointwise engine needs next; the PS input packer and later blocks use this same packaging, so data never needs reshuffling between stages. The one exception is after the final block: the complete $90\times160\times64$ latent is written to memory and read back through the pointwise engine's own input path for VQ, so ordinary convolution and VQ compete for that same resource and cannot run at the same time; this contention is captured by the latency model below.

\subsection{Compute--Communication Latency Model}
\label{sec:arch:model}

\begin{algorithm}[t]
\caption{Compute--Communication Latency Oracle. \textsc{BlockLatency} returns the bottleneck latency among competing read, depthwise, pointwise, and write demands; \textsc{GroupLatency}, which models the pointwise scheduler, is additionally reused for vector quantization.}
\label{alg:latency}
\footnotesize
\begin{algorithmic}[1]
\Statex \textbf{Given:} accelerator $\mathcal{H}(L,Q)$; block workload $\mathcal{W}_j$ as in Eq.~\eqref{eq:block_workload_descriptor}; latent sub-vector dimension $D_m=D/M$ over $N_{\mathrm{pos}}$ positions; evaluated VQ configurations satisfy $K\bmod Q=0$.
\Statex \textbf{Constants} (deterministic cycle counts of the RTL scheduler; each a named RTL construct, none fitted):
\Statex \quad $L=8$,\; $W_{\mathrm{DMA}}=8$ B/cyc,\; $R_{\mathrm{sh}}=2$
\Statex \quad $\delta_{\mathrm{acc}}=6$,\; $\delta_{\mathrm{ppu}}=2$,\; $\delta_{\mathrm{tr}}=1$,\; $\delta_{\mathrm{flush}}=1$,\; $\delta_{\mathrm{row}}=4$
\Statex
\Function{GroupLatency}{$c_{\mathrm{in}},c_{\mathrm{out}},\mathcal{H}$}
  \State $B \gets \lceil c_{\mathrm{out}}/Q \rceil$;\quad $Q_{\mathrm{last}} \gets c_{\mathrm{out}}-(B-1)Q$
  \State $\Gamma \gets c_{\mathrm{in}}+\lceil Q_{\mathrm{last}}/R_{\mathrm{sh}}\rceil+\delta_{\mathrm{acc}}$
  \If{$B=1$}
    \State \Return $\max(c_{\mathrm{out}}+\delta_{\mathrm{ppu}},\,\Gamma)$
    \Comment{single-batch path}
  \EndIf
  \State $\Delta \gets \max(c_{\mathrm{in}}+\delta_{\mathrm{acc}},\,Q+\delta_{\mathrm{ppu}})$
  \State $T_{\mathrm{ret}} \gets c_{\mathrm{out}}+B\delta_{\mathrm{ppu}}$
  \Comment{PPU retirement}
  \State $T_{\mathrm{acc}} \gets c_{\mathrm{in}}+\delta_{\mathrm{acc}}+(B-2)\Delta+\max(\Delta+\delta_{\mathrm{tr}},\Gamma)$
  \Comment{accumulation}
  \State \Return $\max(T_{\mathrm{ret}},T_{\mathrm{acc}})$
\EndFunction
\Statex
\Function{BlockLatency}{$\mathcal{W}_j,\mathcal{H}$}
  \State $P_{\mathrm{in},j} \gets H_jW_j$;\quad $P_j \gets P_{\mathrm{in},j}/s_j^2$;\quad $G_j \gets \lceil W_j/L \rceil$
  \State $T_{\mathrm{read}} \gets P_{\mathrm{in},j}c_{\mathrm{in},j}/W_{\mathrm{DMA}}$
  \State $T_{\mathrm{DW}} \gets (H_j+1)\,[\,c_{\mathrm{in},j}(G_j+\delta_{\mathrm{flush}})+\delta_{\mathrm{row}}\,]$
  \State $T_{\mathrm{PW}} \gets (P_j/L)\cdot\Call{GroupLatency}{c_{\mathrm{in},j},c_{\mathrm{out},j},\mathcal{H}}$
  \State $T_{\mathrm{write}} \gets P_jc_{\mathrm{out},j}/W_{\mathrm{DMA}}$
  \State \Return $\max(T_{\mathrm{read}},T_{\mathrm{DW}},T_{\mathrm{PW}},T_{\mathrm{write}})$
  \Comment{bottleneck latency}
\EndFunction
\Statex
\Function{$T_{\mathrm{enc}}$}{$C,\mathcal{H}$}
  \Comment{analysis-transform latency, Eq.~\eqref{eq:total_latency}}
  \State \Return $\sum_{j=1}^{|C|}\Call{BlockLatency}{\mathcal{W}_j(C),\mathcal{H}}$
\EndFunction
\Statex
\Function{$T_{\mathrm{VQ}}$}{$M,K,\mathcal{H}$}
  \Comment{VQ reuses the same pointwise scheduler}
  \State \Return $(N_{\mathrm{pos}}/L)\cdot\Call{GroupLatency}{D_m,\,MK,\,\mathcal{H}}$
\EndFunction
\end{algorithmic}
\end{algorithm}

For a candidate encoder $C$ running on accelerator $\mathcal{H}(L,Q)$, each block's timing is modeled as the competition between its read, depthwise, pointwise, and write costs, determined by that block's own tensor dimensions. The expressions below come from directly counting cycles in the real RTL design, not from fitting a formula to measured timing.

For block $j$, because only $Q$ output channels can be accumulated in parallel, the $c_{\mathrm{out},j}$ outputs are processed in $B_j=\lceil c_{\mathrm{out},j}/Q\rceil$ batches, with the final batch having valid width $Q_{\mathrm{last},j}=c_{\mathrm{out},j}-(B_j-1)Q$. We further define
\begin{equation}
\begin{aligned}
\Delta_j&=\max\!\left(c_{\mathrm{in},j}+\delta_{\mathrm{acc}},Q+\delta_{\mathrm{ppu}}\right),\\
\Gamma_j&=c_{\mathrm{in},j}+\left\lceil\frac{Q_{\mathrm{last},j}}{R_{\mathrm{sh}}}\right\rceil+\delta_{\mathrm{acc}}.
\end{aligned}
\label{eq:pw_batching}
\end{equation}
Here, $\delta_{\mathrm{acc}}=6$ is the RTL's accumulation cycle count, $\delta_{\mathrm{ppu}}=2$ is the cycles the $L$-lane PPU needs for a full batch, and $R_{\mathrm{sh}}=2$ is the shadow-storage transfer rate in output channels per cycle; $\Delta_j$ is therefore the slower of the accumulation and retirement recurrences, while $\Gamma_j$ captures final accumulation and shadow transfer. For two or more batches, the RTL schedule also exposes a fixed one-cycle control transition, $\delta_{\mathrm{tr}}=1$, during final-batch handling. \textsc{GroupLatency} (Algorithm~\ref{alg:latency}) implements this: when all outputs fit in one batch, the group latency is simply the larger of the PPU's readout time and the accumulation time. When multiple batches are needed, retirement and accumulation instead run as two parallel, overlapping processes, so the latency is set by whichever finishes last, not by their sum. We denote this group latency by $T_{\mathrm{group},j}=\textsc{GroupLatency}(c_{\mathrm{in},j},c_{\mathrm{out},j},\mathcal{H})$.
The corresponding steady-state pointwise latency, a cycle count like all latency quantities in this subsection, is
\begin{equation}
T_{\mathrm{PW},j}=\frac{P_j}{L}T_{\mathrm{group},j}.
\label{eq:pw_latency}
\end{equation}
Here, $P_{\mathrm{in},j}=H_jW_j$ is the block's total input spatial positions, and $P_j=P_{\mathrm{in},j}/s_j^2$ is the number of output spatial positions the pointwise engine processes once the stride has downsampled the input by $s_j$ in each direction. Because activations are INT8 and the implemented AXI4-Stream datapath is 64 bits wide, $W_{\mathrm{DMA}}=8$ activation bytes are transferred per cycle. The remaining latencies are
\begin{equation}
\begin{aligned}
T_{\mathrm{read},j}&=\frac{P_{\mathrm{in},j}c_{\mathrm{in},j}}{W_{\mathrm{DMA}}},\\
T_{\mathrm{DW},j}&=(H_j+1)\left[c_{\mathrm{in},j}\left(G_j+\delta_{\mathrm{flush}}\right)+\delta_{\mathrm{row}}\right],\\
T_{\mathrm{write},j}&=\frac{P_jc_{\mathrm{out},j}}{W_{\mathrm{DMA}}},
\end{aligned}
\label{eq:block_latencies}
\end{equation}
where $G_j=\lceil W_j/L\rceil$ is the number of spatial input groups per row.
The sliding-window generator adds two small fixed delays: $\delta_{\mathrm{flush}}=1$ cycle after each channel's row to flush it out, and $\delta_{\mathrm{row}}=4$ cycles to drain each spatial row. It also runs one extra, input-free row at the very end to flush the last row through, giving $H_j+1$ row passes in total. The depthwise PPU keeps pace with one $L$-activation group per cycle, so it adds no delay of its own; $T_{\mathrm{DW},j}$ therefore reflects this streaming schedule rather than the depthwise layer's raw MAC count.
\begin{equation}
T_{\mathrm{block},j}(C,\mathcal{H})=\max\!\left(T_{\mathrm{read},j},T_{\mathrm{DW},j},T_{\mathrm{PW},j},T_{\mathrm{write},j}\right).
\label{eq:block_time_model}
\end{equation}
$T_{\mathrm{block},j}$, the block's overall latency, is set by whichever of these four costs is largest, its bottleneck. This determines whether adding more arithmetic parallelism can actually reduce block latency: once a different cost dominates, increasing pointwise capacity no longer reduces $T_{\mathrm{block},j}$. Because the encoder blocks execute sequentially with inter-block outputs materialized in DDR, no cross-block compute overlap is assumed, and the total analysis-transform latency is
\begin{equation}
T_{\mathrm{enc}}(C,\mathcal{H})=\sum_{j=1}^{N}T_{\mathrm{block},j}(C,\mathcal{H}).
\label{eq:total_latency}
\end{equation}
Finite startup and termination contribute at most 91 cycles over the deployed blocks ($<0.2\%$ of pointwise latency) and are omitted, together with host turnaround, DMA-launch overhead, per-frame codebook reloads, and transient FIFO effects.

Because VQ mode follows the same pointwise load--accumulate--drain schedule as ordinary convolution, its timing is obtained by mapping the quantizer geometry onto an equivalent pointwise workload, rather than deriving a separate accelerator model. In all VQ configurations evaluated in this work, $K$ is an integer multiple of $Q$, so each sub-codebook occupies an integer number of $Q$-wide output batches, and the total number of such batches per latent position is $B_{\mathrm{VQ}}(M,K,Q)=MK/Q$.

Because sub-codebook boundaries always line up exactly with batch boundaries, every batch scores candidates from just one sub-codebook, using that sub-codebook's own $D_m$-dimensional codewords in full, with no wasted, zero-padded slots. This lets the same pointwise scheduling formula be reused for VQ by substituting $D_m$ (the codeword dimension) in place of the channel count $c_{\mathrm{in}}$, and $MK$ (the total number of codewords across all sub-codebooks) in place of the channel count $c_{\mathrm{out}}$, even though these VQ quantities are not literally channel counts. Substituting these dimensions into the same pointwise scheduler, and taking $N_{\mathrm{pos}}=90\times160$ as the number of spatial positions in the latent, gives the VQ group latency and total latency as
\begin{equation}
\begin{aligned}
T_{\mathrm{group,VQ}}&=\textsc{GroupLatency}(D_m,MK,\mathcal{H}),\\
T_{\mathrm{VQ}}(M,K,\mathcal{H})&=\frac{N_{\mathrm{pos}}}{L}T_{\mathrm{group,VQ}}.
\end{aligned}
\label{eq:vq_latency}
\end{equation}
Reading the latent in and writing out the chosen indices reuse the existing 64-bit streaming ports and finish faster than the pointwise engine's own group latency, so they add no extra time. Likewise, forming the bias and finding the best score happen during the normal accumulator retirement step, adding no separate cycle cost.

The shared-engine formulation exposes a more general quantizer-design rule. Exhaustive codeword scoring requires
\begin{equation}
N_{\mathrm{mul,VQ}}=MKD_m=DK
\label{eq:vq_useful_work}
\end{equation}
useful scalar multiplies per latent position, and the learned codebook stores the same $DK$ scalar values, while the fixed-width index representation needs $M\lceil\log_2 K\rceil$ bits. At fixed $D=64$ and fixed $(M,D_m)=(4,16)$, reducing $K$ therefore decreases arithmetic and codebook storage linearly, and also decreases the number of bits needed per index. This hardware benefit is coupled to rate--distortion behavior: reducing $K$ helps only while the retained codewords still represent the learned latent adequately. We therefore treat $K$ as a post-training deployment variable, with the resulting tradeoff and selection reported in Section~\ref{sec:results:vq}.

For a selected encoder and VQ configuration, the serial PL compute latency is therefore
\begin{equation}
T_{\mathrm{PL,serial}}(C,M,K,\mathcal{H})=
T_{\mathrm{enc}}(C,\mathcal{H})+
T_{\mathrm{VQ}}(M,K,\mathcal{H}),
\label{eq:shared_serial_latency}
\end{equation}
the total time to run analysis and then VQ back to back on the shared engine, since the two cannot execute simultaneously; this sum therefore describes single-frame occupancy rather than sustained frame initiation interval. It also overstates the true minimum: whenever a block is depthwise-bound, the pointwise engine sits idle while depthwise finishes, and an ideal, finer-grained scheduler could fill that idle time with VQ work instead. Doing so would achieve the lower architectural bound $II_{\mathrm{frame}}\ge\sum_{j=1}^{N}T_{\mathrm{PW},j}+T_{\mathrm{VQ}}$ on the frame initiation interval, which counts only actual pointwise occupancy rather than each block's full latency. Reaching this lower bound would require such fine-grained interleaving, whereas the deployed scheduler instead uses coarse-grained analysis and VQ phases. Section~\ref{sec:results:system} reports both the modeled bound and measured initiation interval for the selected implementation.

\subsection{Reference-Workload Hardware Provisioning}
\label{sec:arch:selection}
How large to build a compute engine depends on how much work the workload actually demands from it. We therefore use the original encoder $C_{\mathrm{ref}}$ as a representative workload for sizing the hardware, without specializing the accelerator to that one network: the same reusable depthwise--pointwise engine supports any channel dimensions and tensor shapes at runtime, and the latency equations are validated over a much broader set of configurations. This reference workload is used only to find the practical balance between depthwise, pointwise, and communication costs on the target device.

Doubling $L$ from 8 to 16 would double the datapath from 64 to 128 bits, a full interface redesign rather than a simple parallelism tweak. Holding $Q=32$ fixed, it would also double the structural pointwise DSP requirement from $LQ/2=128$ to 256 DSP48E1 slices, exceeding the device's 220-slice budget before depthwise and post-processing resources are even counted. We therefore hold $L=8$ fixed and instead search over pointwise parallelism, $Q\in\{8,16,32\}$, choosing
\begin{equation}
Q^{\star}\in\underset{Q\in\{8,16,32\}}{\arg\min}\;T_{\mathrm{enc}}\left(C_{\mathrm{ref}},\mathcal{H}(L=8,Q)\right),
\label{eq:q_selection}
\end{equation}
subject to the device's resource limits, which gives $Q^{\star}=32$. Eq.~\eqref{eq:pw_dsp_cost} gives the DSP count for the packed multiplier array itself, but complete DSP usage also depends on how synthesis maps other arithmetic to DSP48E1 versus LUT resources, so we do not extend this formula to configurations we have not actually synthesized.

\subsection{Hardware Cost Oracles and Quantizer Selection}
\label{sec:arch:oracle}

After sizing the reusable accelerator, we fix $\mathcal{H}^\star=\mathcal{H}(L=8,Q=32)$. For candidate encoder $C_i$, substituting its tensor geometry into Eqs.~\eqref{eq:pw_batching}--\eqref{eq:total_latency} gives the platform-specific cost
\begin{equation}
T_{\mathrm{HW}}(C_i)=T_{\mathrm{enc}}\!\left(C_i,\mathcal{H}^\star\right),
\label{eq:hardware_cost_oracle}
\end{equation}
which requires no candidate-specific training or hardware implementation to evaluate. Section~\ref{sec:method:nas} describes how this oracle guides encoder selection. The same approach is reused for VQ: with $(M,D_m)=(4,16)$ already fixed, the model predicts hardware speed for each candidate $K$, while separate experiments measure reconstruction quality and entropy-coded rate for that same $K$. Section~\ref{sec:method:vq_sweep} combines predicted speed with measured quality and rate to choose the deployed cardinality.

\section{Experimental Methodology}
\label{sec:method}

With the latency model and accelerator design from Section~\ref{sec:arch} fixed, this section validates that model against silicon before using it to run the encoder search and VQ refinement, and closes with the evaluation protocol used to report results.

\subsection{Latency Model Validation}
\label{sec:method:hardware_validation}

The accelerator is implemented in SystemVerilog on an XC7Z020CLG484-1 Zynq-7020, using Vivado 2020.2 at 100~MHz. The timed edge path covers group-major input packing, the depthwise--pointwise analysis transform, shared-pointwise VQ, and PS rANS coding, and excludes image acquisition and the cloud decoder. Board timings come from the ARM Global Timer after warm-up.

We report modeled and measured analysis and VQ latency. The pointwise scheduler is checked against 100 RTL configurations spanning $Q\in\{8,16,32\}$ and varying channel geometries, while the depthwise scheduler is checked against 100 configurations across strides, channel counts, spatial dimensions, and ragged widths; both are then compared against Zynq-7020 measurements.

Hardware ablations compare serialized versus split-shadow pointwise retirement, and fused versus DDR-materialized depthwise--pointwise execution. Shared-pointwise VQ is checked against the software quantizer on random, exact-tie, and INT8-extreme inputs, then verified index-for-index across full frames.

\subsection{Encoder Exploration}
\label{sec:method:nas}

With the accelerator $\mathcal{H}^\star=\mathcal{H}(L=8,Q=32)$ validated above, we rank candidate encoder schedules $C=(c_1,\ldots,c_N)$ by modeled latency $T_{\mathrm{HW}}$ (Eq.~\eqref{eq:hardware_cost_oracle}) before training any of them. Because proxy training dominates exploration cost, candidates are trained at two matched modeled-latency levels: that of $C_{\mathrm{ref}}$ ($\approx$17.7~ms) and a faster level near 13.4--14.0~ms. Search~A holds the six-block depth of $C_{\mathrm{ref}}$ and redistributes channel width between the early, high-resolution blocks and the later, low-resolution ones; Search~B varies depth from three to six blocks. Depth and channel placement are thus compared at matched modeled latency. All candidates use the reference VQ setting $(M_0,K_0,D_{m,0})=(4,256,16)$, so VQ does not affect ranking.

After identical short proxy training, we compare candidates on image quality (MS-SSIM and PSNR) and $T_{\mathrm{HW}}$; since every candidate shares the same rate preset, rate is not a selection criterion. The winning topology is then compared with $C_{\mathrm{ref}}$ and trained as prescribed in Section~\ref{sec:methodology:training}.

\subsection{VQ Codebook Refinement}
\label{sec:method:vq_sweep}

With $C^\star=(16,48,64)$ and $\mathcal{H}^\star$ fixed, we independently refine the codebook cardinality for each of the three rate presets over $K\in\{256,128,96,64,32\}$, holding $M=4$ and $D_m=16$ constant. For each $K$, we rank each sub-codebook's original 256 entries by how often they were used during training (their EMA usage estimate) and keep only the top $K$. Each candidate $K$ is then scored on reconstruction quality, entropy-coded rate, codebook storage, and modeled VQ latency $T_{\mathrm{VQ}}$ (Eq.~\eqref{eq:vq_latency}). Section~\ref{sec:results:vq} reports the findings and the selected cardinality.

\subsection{Evaluation Metrics and Protocol}
\label{sec:method:eval}

For the finalized codec, we evaluate reconstruction quality with two traditional distortion metrics, PSNR and MS-SSIM, and two perceptual metrics, LPIPS and BRISQUE. PSNR is reported for comparability with prior FPGA LIC work, though it correlates poorly with perceived visual quality. MS-SSIM captures structural similarity. LPIPS is a learned perceptual metric, while BRISQUE is a reference-free perceptual quality metric. The coded rate is $R_{\mathrm{coded}}=8N_{\mathrm{bytes}}/(H_{\mathrm{img}}W_{\mathrm{img}})$~bpp, where $N_{\mathrm{bytes}}$ is the total encoded payload plus header, in bytes. Rate--distortion is evaluated at native resolution on the CLIC~2017 and Kodak test suites.

For hardware efficiency, we report useful convolutional arithmetic per block as $N_{\mathrm{MAC},j}=P_j(9c_{\mathrm{in},j}+c_{\mathrm{in},j}c_{\mathrm{out},j})$, the $9c_{\mathrm{in},j}$ term from the $3\times3$ depthwise filter and the $c_{\mathrm{in},j}c_{\mathrm{out},j}$ term from the pointwise stage; totals are reported per frame, and, where noted, normalized by input pixel count, separately from modeled shared-engine latency and VQ work.

\section{Experimental Results and Co-Design Analysis}
\label{sec:results}

The evaluation follows the staged co-design developed above. We first validate the compute--communication model and the provisioned accelerator, then evaluate how the resulting latency oracle supports target-platform-aware construction and selection of encoder candidates and whether its predicted-latency ordering differs from useful-MAC ordering. We next evaluate post-training codebook-cardinality refinement and the resulting analysis--VQ workload on the shared pointwise engine, before reporting shared-engine system behavior and rate--distortion performance.

\subsection{Hardware-Model Validation and Encoder Selection}
\label{sec:results:validation}

\begin{table}[!t]\caption{Per-Block Latency-Model Validation of the $16$--$48$--$64$ Transform}\label{tab:deployed_block_validation}
\centering
\footnotesize
\setlength{\tabcolsep}{4.5pt}
\begin{adjustbox}{max width=\linewidth}
\begin{tabular}{@{}lrrrr@{}}\toprule Block & Predicted (cyc) & Measured (cyc) & Error & Bottleneck \\\midrule 1 & 518,400 & 519,904 & $-0.289\%$ & PW \\2 & 469,300 & 470,279 & $-0.208\%$ & DW \\3 & 356,932 & 357,990 & $-0.296\%$ & DW \\\midrule Analysis total & 1,344,632 & 1,348,173 & $-0.263\%$ & -- \\VQ & 489,600 & 489,840 & $-0.049\%$ & PW \\\midrule Combined & 1,834,232 & 1,838,013 & $-0.206\%$ & -- \\\bottomrule
\end{tabular}
\end{adjustbox}
\end{table}

Across more than one hundred pointwise RTL tensor configurations spanning $Q\in\{8,16,32\}$, the analytical schedule reproduces implemented cycle counts to within 0.05 cycle/group, while independent pointwise-bound measurements agree with model predictions to approximately 1\%. The depthwise schedule was independently verified over 100 RTL configurations spanning channel counts, image widths, ragged widths, and both supported strides; its error is a small, constant offset from pipeline fill and drain, confirming that the model correctly captures the ongoing per-cycle rate. Table~\ref{tab:deployed_block_validation} provides an independent silicon check on the final $16$--$48$--$64$ transform, for which the aggregate prediction differs from measurement by only 0.264\% and the model identifies the bottleneck correctly in every block.

\begin{figure}[!t]\centering\includegraphics[width=\linewidth]{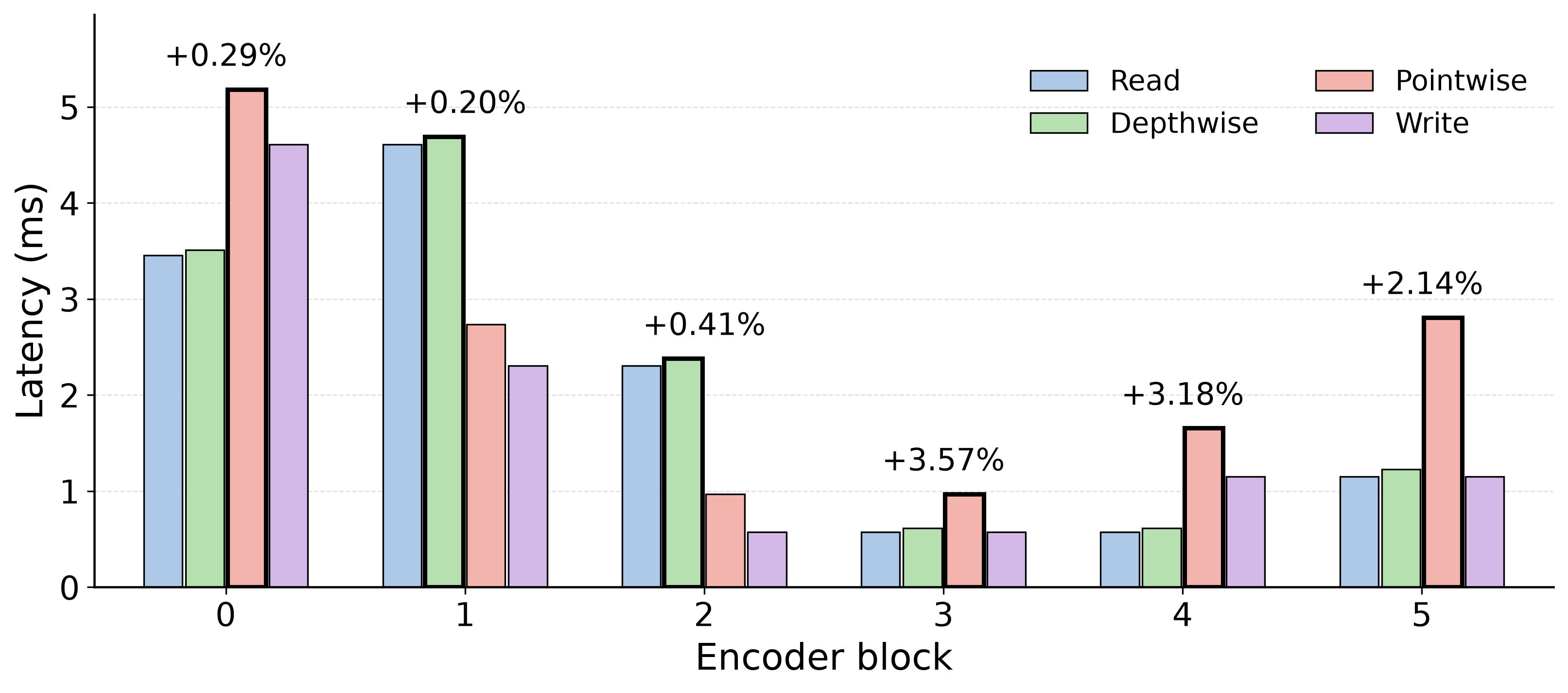}
\caption{Latency-model validation on the six-block reference workload $C_{\mathrm{ref}}$ at $L=8,Q=32$. Bars show the modeled read, depthwise, pointwise, and write latencies per block. The bold bar marks each block's bottleneck, and the percentage above it is the amount by which the measured block latency exceeds the predicted one.}
\label{fig:latency}
\end{figure}

Fig.~\ref{fig:latency} further shows that the model tracks changes in the bottleneck across the six-block reference workload. Hardware provisioning on this workload selects $L=8,Q=32$: increasing pointwise parallelism from $Q=16$ to $Q=32$ and introducing split-shadow accumulator retirement together reduce measured PL/DMA latency by 35.6\%, while depthwise--pointwise fusion reduces external activation traffic from 26.266 to 18.432~MB/frame. These measurements establish the reusable accelerator configuration on which the latency model is frozen before encoder exploration.

Table~\ref{tab:nas_results} reports the structured encoder candidates evaluated under the same short proxy-training budget (Section~\ref{sec:method:nas}) using this fixed latency oracle. The predicted-latency ordering is not equivalent to useful-MAC ordering: the six-block $16$--$16$--$16$--$16$--$32$--$64$ candidate has 21.1\% fewer useful MACs than the selected $16$--$48$--$64$ encoder, yet higher modeled latency. Thus, arithmetic count alone does not preserve latency ordering when the bottleneck changes with tensor geometry.

Among the evaluated candidates, $16$--$48$--$64$ simultaneously achieves the highest proxy PSNR and lowest modeled hardware latency and is therefore selected. Relative to the six-block $16$--$32$--$32$--$32$--$64$--$64$ reference, it reduces convolutional weights from 10,363 to 4,491 (56.7\%), modeled latency by 24.0\%, and modeled external activation traffic from 18.432 to 16.589~MB/frame (10.0\%).

\begin{table}[!t]
\caption{Hardware-Model-Guided Encoder Exploration Results}
\label{tab:nas_results}
\centering
%\scriptsize
\setlength{\tabcolsep}{6pt}
\begin{adjustbox}{max width=\linewidth}\begin{tabular}{@{}lrrrr@{}}
\toprule 
Encoder &$\sum_j N_{\mathrm{MAC},j}$ &$T_{\mathrm{HW}}$&PSNR &MS-SSIM \\&(MMAC/frame) &(ms) &(dB) &(dB) \\
\midrule\textbf{16--48--64}& \textbf{120.269}& \textbf{13.446}& \textbf{23.405}& \textbf{9.475} \\16--16--64--64& 124.416& 13.909& 22.394& 7.925 \\16--64--64--64& 219.110& 17.645& 22.727& 8.465 \\16--16--48--32--64& 115.430& 13.988& 21.927& 7.142 \\16--48--48--48--64& 199.066& 17.766& 22.283& 7.890 \\16--32--32--32--64--64& 193.766& 17.695& 21.768& 7.119 \\16--32--48--64--32--64& 203.213& 17.695& 21.708& 7.420 \\16--16--16--16--32--64& 94.925& 13.952& 20.957& 6.245 \\
\bottomrule
\end{tabular}
\end{adjustbox}
\end{table}

\subsection{Codebook Refinement and Shared-Engine Selection}
\label{sec:results:vq}

With the encoder topology $C^\star=(16,48,64)$ fixed, assignment statistics from the trained quantizers show substantial codebook under-utilization, motivating post-training cardinality refinement. Table~\ref{tab:vq_refinement} reports the resulting latency--storage--quality tradeoff for the mid-rate preset while holding $M=4$ and $D_m=16$ fixed.

\begin{table}[!t]
\caption{Representative Post-Training Codebook-Cardinality Refinement for the Mid-Rate Preset}
\label{tab:vq_refinement}
\centering
\resizebox{\columnwidth}{!}{\begin{tabular}{@{}rrrrrrr@{}}
\toprule
$K$ & \makecell{Uncoded\\index bits} & \makecell{PSNR\\(dB)} & \makecell{MS-SSIM\\(dB)} & \makecell{Codebook\\(KiB)} & \makecell{Useful\\mult./pos.} & \makecell{$T_{\mathrm{VQ}}$\\(ms)} \\
\midrule
256 &32 &29.29 &13.38 &16 &16,384 &19.584\\
128 &28 &29.21 &13.33 &8 &8,192 &9.792 \\
96 &28 &29.10 &13.28 &6 &6,144 &7.344 \\
\textbf{64} &\textbf{24} &\textbf{28.69} &\textbf{13.06} &\textbf{4} &\textbf{4,096} &\textbf{4.896} \\
32 &20 &27.39 &12.35 &2 &2,048 &2.448 \\
\bottomrule
\end{tabular}}
\end{table}

Across all three independently trained rate presets, reducing $K$ produced the same qualitative quality trend, with a clear quality knee at $K=64$. Table~\ref{tab:vq_refinement} reports the mid-rate preset as a representative sweep. Reducing $K$ from 256 to 64 cuts exhaustive-search arithmetic, codebook storage, and modeled shared-engine VQ latency by roughly $4\times$, while retaining most of the reconstruction quality; reducing further to $K=32$ produces a substantially larger quality loss. We therefore select $K^\star=64$ for all three rate presets.

We then re-evaluate pointwise width using this final analysis--VQ workload. Increasing to an analytical $Q=64$ improves combined modeled analysis--VQ latency by under 1\% while exceeding the implemented device's pointwise DSP budget, so $Q^\star=32$ remains the selected operating point.

\begin{table}[!t]
\caption{Resource Utilization of the Final Accelerator}
\label{tab:final_utilization}
\centering
\footnotesize\setlength{\tabcolsep}{4.5pt}
\begin{tabular}{@{}lrrr@{}}
\toprule Resource &Used &Available &Utilization \\\midrule Slice LUTs &33,994 &53,200 &63.90\% \\Slice registers &32,353 &106,400 &30.41\% \\Block RAM tiles &82.5 &140 &58.93\% \\DSP48E1 &220 &220 &100.00\% \\
\bottomrule
\end{tabular}
\end{table}

\begin{figure}[!t]
\centering
\includegraphics[width=\columnwidth]{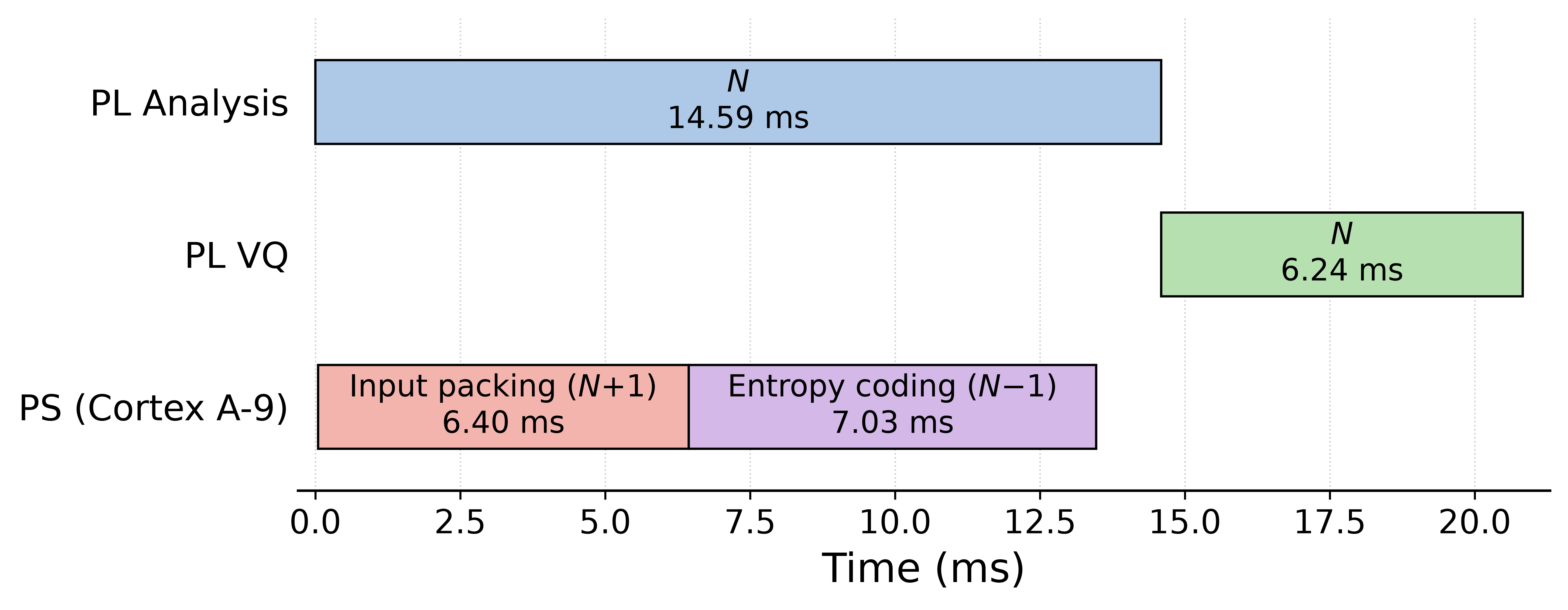}
\caption{Measured multi-frame execution schedule. Input packing for frame $N{+}1$ and PS (Cortex-A9) rANS coding for frame $N{-}1$ overlap with the analysis stage of frame $N$, followed by the VQ stage for that same frame, so a new frame starts every 20.84~ms rather than after the sum of all four stage latencies. Stage-level timings include associated PS-side control and data-movement overheads, whereas Table~\ref{tab:deployed_block_validation} reports accelerator-only PL cycle counts used for latency-model validation.}
\label{fig:pipeline_schedule}
\end{figure}

\subsection{Final FPGA Implementation and System Performance}
\label{sec:results:system}

Post-place utilization of the final accelerator is summarized in Table~\ref{tab:final_utilization}. The implementation uses all 220 DSP48E1 slices, leaving 36.1\% of LUTs, 69.6\% of registers, and 41.1\% of BRAM capacity available for control and system integration.

Table~\ref{tab:deployed_block_validation} confirms the latency model also remains accurate for VQ and for the combined analysis--VQ workload. The deployed implementation additionally reloads the VQ codebook into the shared weight memory once per frame, adding a measured 0.9599~ms, giving a measured serial analysis--reload--VQ latency of 19.3400~ms.

As shown in Fig.~\ref{fig:pipeline_schedule}, host-side work is overlapped across adjacent frames rather than added serially to PL execution. Under this schedule, the complete edge pipeline sustains a measured frame initiation interval of 20.8405~ms, corresponding to 47.98~frames/s at 720p. PS entropy coding, including index unpacking, rANS coding, and payload assembly, requires 5.3730~ms of CPU time but is 99.9\% overlapped with PL analysis, leaving under 0.004~ms combined exposed; host-side entropy coding therefore does not determine steady-state throughput.

PMBus measurements give $2.0567\pm0.0053$~W of monitored board-rail power. An independent power-measurement loop reports a 20.8287-ms frame period, agreeing with the timing measurement to within $0.06\%$, and yields an energy cost of 42.84~mJ/frame. These values are regulator-output measurements and exclude regulator conversion losses and the unmonitored 5-V USB rail.

Finally, the shared-pointwise latency model gives a 16.416-ms ideal compute-only initiation-interval floor, compared with 18.342~ms for coarse-grained analysis--VQ execution, leaving roughly 10.5\% of compute-level scheduling headroom unexploited by the deployed coarse-grained schedule.

\begin{figure*}[!t]
\centering
\includegraphics[width=\linewidth]{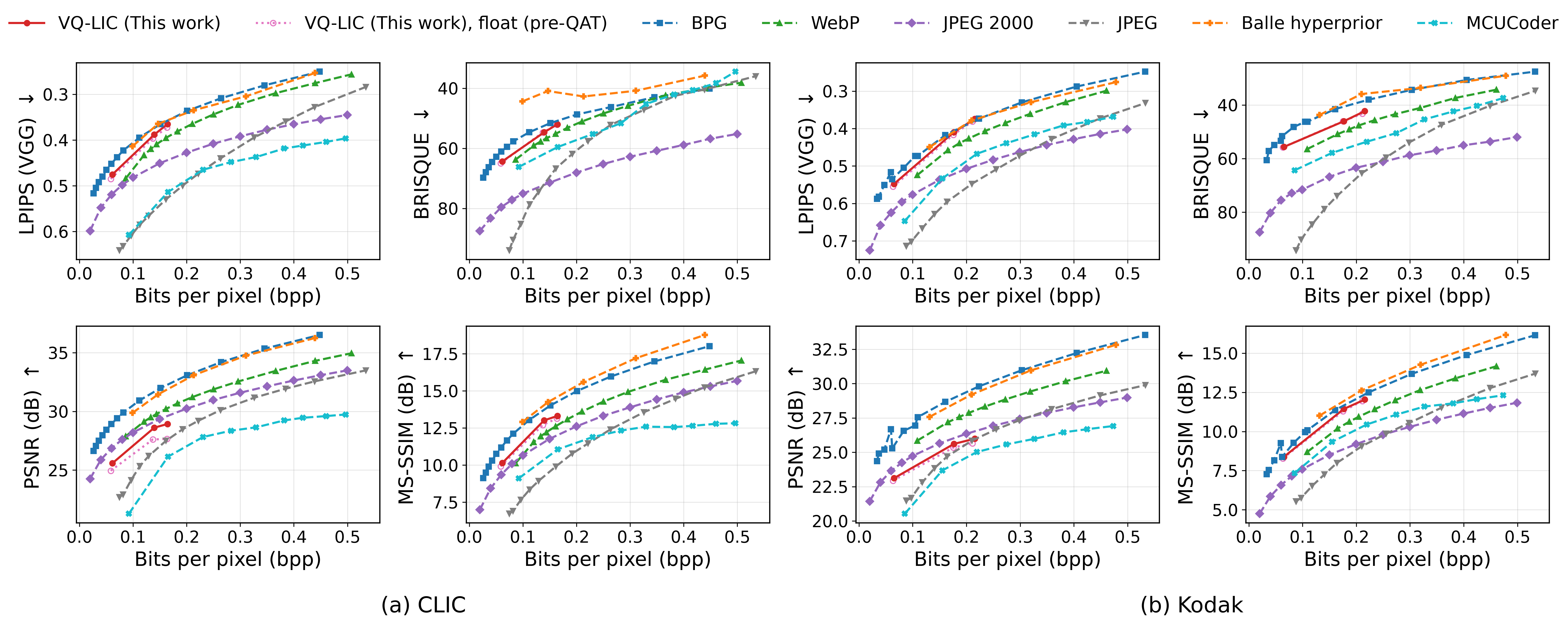}
\caption{Rate--distortion and perceptual-quality comparison of VQ-LIC (three independently trained rate presets, shown for the final INT8/QAT and pre-QAT floating-point models) against BPG, WebP, JPEG~2000, JPEG, the Ball\'e hyperprior codec, and MCUCoder, on CLIC~2017 and Kodak at native resolution. Metrics shown are LPIPS, BRISQUE, PSNR, and MS-SSIM; arrows indicate whether lower ($\downarrow$) or higher ($\uparrow$) is better.}
\label{fig:rd}
\end{figure*}

\begin{figure*}[!t]
\centering
\includegraphics[width=\textwidth]{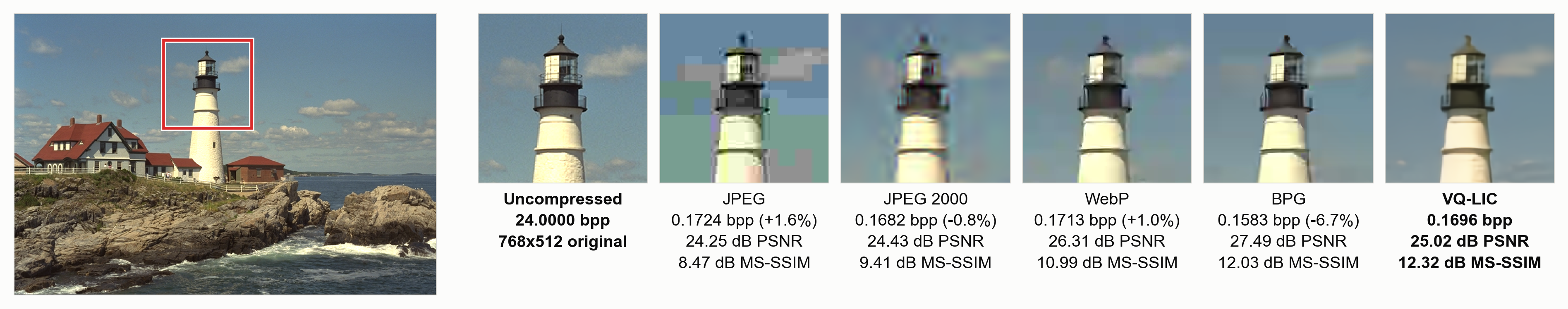}
\caption{Rate--distortion and perceptual-quality comparison of VQ-LIC (mid-rate preset, final INT8/QAT model) against traditional baselines.}
\label{fig:parrot}
\end{figure*}

\begin{table}[t]
\centering
\caption{Deployment comparison between VQ-LIC and a commercial edge-accelerator platform.}
\label{tab:coral_comparison}
\resizebox{\columnwidth}{!}{
\begin{tabular}{lcc}
\hline
\textbf{Metric} &
\makecell{\textbf{VQ-LIC}\\\textbf{(Zynq-7020)}} &
\makecell{\textbf{Coral TPU +}\\\textbf{Raspberry Pi 5}} \\
\hline
Analysis-transform latency (ms) & 13.48 & 24.80 \\
Complete edge frame interval (ms) & 20.84 & 27.80 \\
Throughput (frame/s) & 47.98 & 35.97 \\
Platform power (W) & 2.06$^{a}$ & 7.80$^{b}$ \\
Energy per frame (mJ) & 42.84$^{a}$ & 216.84$^{b}$ \\
VQ execution & PL PW engine & RPi 5 \\
Entropy coding (rANS) & Cortex-A9 & RPi 5 \\
\hline
\end{tabular}
}
\begin{minipage}{\columnwidth}
\footnotesize
$^{a}$ Measured at the monitored regulator outputs, excluding regulator conversion losses and the unmonitored 5-V USB rail.
$^{b}$ Coral I/O and invocation (24.80~ms), Pi 5 VQ (1.20~ms), and rANS (1.80~ms) are measured latencies, summed in series to give the 27.80~ms frame interval. Power (7.80~W) is the gross system power draw.
\end{minipage}
\end{table}

\subsection{Rate--Distortion Performance}
\label{sec:results:rd}

Fig.~\ref{fig:rd} evaluates VQ-LIC on unseen CLIC~2017 validation and Kodak images at their native resolutions across three independently trained low-, mid-, and high-rate presets. VQ-LIC closes much of the rate--distortion gap on structural and perceptual metrics. It outperforms JPEG, JPEG~2000, and WebP on MS-SSIM, LPIPS, and BRISQUE, which fits the structurally weighted training objective in Section~\ref{sec:methodology:decoder}, and it also outright outperforms the similarly sized MCUCoder across all four metrics, while narrowing, though not closing, the gap to the far larger Ball\'e hyperprior codec on the perceptual metrics.

Fig.~\ref{fig:parrot} illustrates this qualitatively on a representative image, where VQ-LIC's reconstruction preserves more perceptible structure than the conventional codecs at a matched bitrate. BPG still leads VQ-LIC in raw PSNR. However, it is worth noting that BPG also leads Ball\'e et al.'s hyperprior codec~\cite{balle2018}, which uses a 5.08M-parameter encoder, $1{,}130\times$ larger than VQ-LIC's, and still falls short of BPG's PSNR by 0.15--0.35~dB; even a far larger neural codec pays a real compute cost to get near BPG. On the other end, MCUCoder's encoder has about 10.5K parameters, close to VQ-LIC's own size, yet VQ-LIC outperforms it in PSNR by 0.97--2.59~dB at equal or lower bpp every time. VQ-LIC therefore outperforms a similarly sized encoder (MCUCoder) while reaching a rate--distortion range similar to a model three orders of magnitude larger (the Ball\'e hyperprior codec), running at just 0.1945~kMAC/pixel on a severely resource-constrained platform.

\subsection{Comparison Against a Commodity Edge Platform}
\label{sec:coral}

To contextualize the benefits of application-specific FPGA co-design, we additionally deployed the analysis transform on a Google Coral Edge TPU with a Raspberry Pi 5 host executing VQ and entropy coding. Table~\ref{tab:coral_comparison} summarizes the comparison. The Zynq-7020 implementation completes the analysis stage nearly $2\times$ faster and sustains roughly $1.3\times$ higher throughput, while drawing under a third of the commercial platform's power and under a fifth of its energy per frame. The comparison illustrates the benefit of tailoring the accelerator beyond convolution alone: VQ scoring executes on the same programmable pointwise datapath used by the analysis transform, avoiding a separate host-side quantization stage that the commodity platform requires.

\begin{table*}[!t]
\caption{Comparison of 720p FPGA Learned-Image-Compression Implementations}
\label{tab:priorwork}
\centering
\setlength{\tabcolsep}{3.5pt}
\begin{adjustbox}{max width=\textwidth}
\begin{tabular}{@{}lcccccccccccc@{}}
\toprule
Work &
Platform &
Precision &
\makecell{Clock\\(MHz)} &
LUTs &
DSPs &
BRAM &
\makecell{Throughput\\(fps)} &
\makecell{Energy/frame\\(J)} &
\makecell{PSNR$^{d}$\\(dB)} &
\makecell{MS-SSIM$^{d}$\\(dB)} &
bpp$^{d}$ &
\makecell{Encoder\\kMAC/px} \\
\midrule

\textbf{VQ-LIC (This work)}
& Zynq-7020
& INT8
& 100
& 34.0k
& 220
& 82.5
& 47.98
& 0.043
& 25.99
& 12.00
& 0.2158
& 0.195 \\

X-LIC~\cite{chen2025}
& ZCU102
& INT8
& 200
& 150.3k
& 1522
& 792
& 32.95
& 0.060
& 27.02$^{a}$
& \NR{}
& \NR{}
& 11.224$^{c}$ \\

StreamLIC~\cite{zhang2025streamlic}
& Alveo U50
& INT12
& 250
& \NR{}
& 2992
& \NR{}
& 41.4$^{b}$
& \NR{}
& \NR{}
& \NR{}
& \NR{} 
& 19.30$^{c}$
\\

Sun et al.~\cite{sun2024jetcas}
& KU115
& INT8
& 200
& \NR{}
& 4160
& \NR{}
& 37.74
& \NR{}
& 30.06
& \NR{}
& 0.2531
& 41.030$^{c}$ \\

F-LIC~\cite{sun2022flic}
& VCU118
& Fix8
& 200
& 339.9k
& 4146
& 672
& 40.69
& 1.410
& 29.94
& \NR{}
& 0.2542
& 40.332$^{c}$ \\

FPX-NIC~\cite{jia2022}
& ZCU104
& INT8
& 350
& \NR{}
& \NR{}
& \NR{}
& 3.90
& 1.680
& 29.89
& 15.09
& 0.2565
& 178.743$^{c}$ \\

\bottomrule
\end{tabular}
\end{adjustbox}

\begin{minipage}{1\textwidth}
\footnotesize
$^{a}$ X-LIC PSNR corresponds to the operating point reported for the 720p implementation.
$^{b}$ StreamLIC throughput is estimated from hardware synthesis and simulation for 720p inputs below 0.5 bpp; a corresponding PSNR--bpp operating point was not available for this configuration.
$^{c}$ Encoder complexity counts neural MACs required along the encoding path, including hyper-analysis/hyper-synthesis or VQ where applicable, and excludes arithmetic/range/rANS coding. Values not directly reported are calculated from the published network or component dimensions.
$^{d}$ PSNR, MS-SSIM, and bpp are reported for Kodak.
\end{minipage}
\end{table*}

\subsection{Comparison with Prior FPGA Accelerators}
\label{sec:comparison}

Table~\ref{tab:priorwork} compares representative FPGA LIC implementations reported at approximately 720p. Since platforms, codecs, and test conditions differ, the values are each source's own reported numbers, meant to show general scale rather than a precise, apples-to-apples comparison.

The principal distinction in Table~\ref{tab:priorwork} is the hardware resource regime. VQ-LIC sustains a measured complete edge-path throughput of 47.98 frames/s using 220 DSP48E1 slices on the Zynq-7020, while StreamLIC and F-LIC report 2992 and 4146 DSPs, respectively; these counts should not be read as efficiency ratios across devices, but they show that real-time learned analysis and vector quantization can run within a much smaller multiplier budget than the other implementations considered here. At a modest PSNR tradeoff, VQ-LIC also reaches a substantially lower bitrate, higher throughput, and lower energy per frame than its closest competitors in reconstruction quality, underscoring that VQ-LIC's resource savings come at only a modest cost to rate--distortion, not speed or energy.

\section{Conclusion}
\label{sec:conclusion}

This work presented VQ-LIC, a staged algorithm--architecture co-design for learned image encoding on a severely resource-constrained FPGA. A validated compute--communication model captures the competing read, depthwise, pointwise, and write latencies of the realized accelerator and is reused as a target-platform latency oracle during encoder exploration. The resulting $16$--$48$--$64$ transform reduces modeled latency by 24.0\% relative to the six-block reference while improving proxy reconstruction quality. The same pointwise datapath is subsequently reused for entropy-constrained VQ codeword scoring, eliminating a dedicated VQ compute engine, and post-training cardinality refinement reduces exhaustive-search arithmetic and codebook storage by $4\times$. With pre-shared conditional rANS tables, the selected codec achieves 0.1398~bpp at 28.69~dB PSNR and 13.06~dB MS-SSIM on the mid-rate CLIC operating point, outperforming a similarly sized neural encoder and reaching a rate--distortion range comparable to a model three orders of magnitude larger. On silicon, measured analysis and VQ latencies agree with the latency model's predictions within 0.26\% and 0.05\%, respectively. The complete 0.1945-kMAC/pixel analysis--VQ workload is realized within all 220 DSP48E1 slices of the Zynq-7020, an order of magnitude fewer than comparable FPGA LIC accelerators, which VQ-LIC also outpaces in bitrate, throughput, and energy per frame at a modest PSNR tradeoff; the deployed 720p edge pipeline sustains 47.98 frames/s at 42.84 mJ/frame.

%\section*{Artifact Availability}
%The machine-learning training and evaluation code, the latency model, and the RTL source code are publicly available at \href{https://github.com/abdullahbinfaisal/VQ-LIC}{\texttt{github.com/abdullahbinfaisal/VQ-LIC}}.

\section*{Acknowledgment}
The authors used a large language model to generate the initial scaffolding of Algorithm~1 and Tables~I--VI, and for grammar and prose-flow editing. All technical content and data are the authors' own or drawn from cited work; the AI system generated no results, claims, or scientific content.

\IfFileExists{IEEEtran.bst}{\bibliographystyle{IEEEtran}}{\bibliographystyle{unsrt}}
\bibliography{references}

\end{document}